\documentclass[10pt,aps,pre,amsmath,amsfonts,amssymb,floatfix,superscriptaddress,notitlepage,nofootinbib]{revtex4}
\usepackage[utf8]{inputenc}
\usepackage{amsmath}
\usepackage{amsfonts}
\usepackage{amssymb}
\usepackage{graphicx}
\usepackage{float}
\usepackage[toc,page]{appendix}
\usepackage{bm}
\usepackage{mathrsfs}
\usepackage{mathtools}
\usepackage{xcolor}

\usepackage[ruled,vlined]{algorithm2e}

\DeclareMathOperator\x{\mathbf{X}}
\DeclareMathOperator\z{\mathbf{Z}}
\DeclareMathOperator\pat{\bm{\xi}}

\DeclareMathOperator\xs{\begin{bmatrix} \x\\ \sigma \end{bmatrix}}
\DeclareMathOperator{\R}{\mathbb{R}}
\DeclareMathOperator\p{p}
\DeclareMathOperator\fe{\mathscr{L}}
\DeclareMathOperator\lm{\mu}
\DeclareMathOperator\Ov{\mathscr{O}}
\DeclareMathOperator\Co{\mathscr{C}}
\DeclareMathOperator{\Smat}{\underline{\underline{S}}}
\DeclareMathOperator\fv{\mathbf{f}}

\def\de{\mathrm d}

\begin{document}

\title{Surfing on minima of isostatic landscapes:
 avalanches and unjamming transition}

\author{Silvio Franz}
\affiliation{Universit\'e Paris-Saclay, CNRS, LPTMS, 91405, Orsay, France}

\author{Antonio Sclocchi}
\affiliation{Universit\'e Paris-Saclay, CNRS, LPTMS, 91405, Orsay, France}

\author{Pierfrancesco Urbani}
\affiliation{Universit\'e Paris-Saclay, CNRS, CEA, Institut de physique th\'eorique, 91191, Gif-sur-Yvette, France.}

\begin{abstract}
Recently, we showed that optimization problems, both in infinite as well as in finite dimensions, for continuous variables
and soft excluded volume constraints, can display entire isostatic phases where local minima of the cost function are marginally stable configurations endowed with non-linear excitations \cite{FSU19, FSU20}.  In this work we describe an athermal adiabatic algorithm to explore with continuity 
the corresponding rough high-dimensional landscape.  We concentrate on a prototype problem of this kind, the spherical perceptron optimization problem with linear cost function (hinge loss).  This algorithm allows to 'surf' between isostatic marginally stable configurations and to investigate some properties of such landscape. In particular we focus on the statistics of avalanches occurring when local minima are destabilized. We show that when perturbing such minima, the system undergoes plastic rearrangements whose size is power law distributed and we characterize the corresponding critical exponent.  Finally we investigate the critical properties of the unjamming transition, showing that the linear interaction potential gives rise to logarithmic behavior in the scaling of energy and pressure as a function of the distance from the unjamming point.  For some quantities, the logarithmic corrections can be gauged out. This is the case of the number of soft constraints that are violated as a function of the distance from jamming which follows a non-trivial power law behavior. 
\end{abstract}

\maketitle

\tableofcontents

\section{Introduction}
Marginally stable minima are of central importance in glassy physics and non-convex optimization problems.
Low temperature relaxation dynamics of infinite-dimensional glassy models fails to reach low energy absolutely 
stable minima and the system remains stuck on a manifold of marginal minima.  An important conjecture is that 
when variables have sufficiently long range interactions, any local optimization algorithm can only reach marginally stable, maybe  
sub-optimal, minima. This is certainly true for gradient descent, simulated annealing \cite{CK93,FFR19}, and has recently 
shown to be the case also for message passing algorithms \cite{EMS20}. While we know that in general not all marginal minima are 
dynamically accessible, the full characterization of the dynamic marginal manifold of high dimensional models is an open problem.  This is for example the case of optimization problems appearing both in spin or structural glass models
where low lying states are marginal but appear to be separated by extensive barriers \cite{Ga85, GKS85, KPUZ13}, suggesting exponential relaxations 
times for generic local algorithms \cite{GS14,GJS19,CGPRM19,GJ19}.

Given that, it is important to understand the properties of marginally stable states in generic random optimization problems and how search algorithms behave when falling into such states. A related and important question is how the landscape of such states changes once it is perturbed in some way \cite{BU16, FS17, JUZY18, SGB20}.

Among marginally stable minima, we can distinguish two classes. On the one hand one has \emph{linear} marginally stable 
configurations which are harmonic minima whose properties are controlled by the Hessian of the cost function in those minima.
Belong to this class for example spherical spin glass models for which a picture of local search algorithms such as gradient descent has emerged with
new interesting spinoffs very recently \cite{CK93,FFR19,FFR20,SKUZ19}.

However there are situations in which either the Hessian of local minima is not well defined, or where it is so singular that the relevant excitations above those minima are non-linear in nature.  Such \emph{non-linear} marginally stable states appear when variables are discrete, e.g. in Ising fully connected spin-glasses \cite{An78,Pa02, BKS08, SYM18}. More importantly, this situation has been shown also to be present in continuous systems at jamming critical points. Originally found in the investigation of random packing of low dimensional spheres \cite{LN98,OLLN02,OSLN03} (see \cite{LN10,LNSW10,He09} for reviews), jamming critical points have been shown to appear in a large class of non-convex constraint satisfaction problems with continuous variables, that includes high-dimensional models such as high-dimensional spheres \cite{CKPUZ14, CKPUZ17, PUZ20},  non-convex neural networks and continuous colouring \cite{FP16,FPSUZ17,Y18, FHU19}.
The jamming point can be reached by an adversarial competition: one defines a cost or energy, function, usually chosen with a 
harmonic or 'Hertzian'  shape \cite{OLLN02,OSLN03}, that penalizes violated constraints. Then one finds the maximum possible number 
of constraints such that the minimum energy is zero.  Therefore jamming is a fine tuned point where the energy 
of the optimization problem is still zero, but no constraints can be added without avoiding some of them becoming unsatisfied.  As soon as the system is compressed beyond jamming, either by increasing the number of constraints or by rendering them more difficult,  most of the critical features of the jamming point disappear. 

This picture has been recently modified in \cite{FSU19,FSU20} where we showed that by properly choosing the cost for constraint violation
the critical properties of the jamming transition survive in an entire critical, self-organized, marginally stable phase. 
In \cite{FSU19} we have shown that in the spherical perceptron optimization  problem with linear cost function (hinge loss) in 
addition to a non-critical jammed phase, there is 
a whole region in the phase diagram which is made of jammed non-linear marginally stable minima with critical properties 
similar to the jamming point.
This has shown that the universality class of jamming does not require fine tuning of the parameters, but can emerge generically in an optimization setting if the cost function is not differentiable. Moreover, in \cite{FSU20} it was found  that such new phase is also present in systems of soft spheres with a linear cost for overlap between the particles.  It goes with this line that one can ask what is the fate of those non-linear minima when perturbations to the system are applied by tilting the cost function and how the corresponding energy landscape is explored by greedy algorithms. Non-linear marginally stable states generically give rise to scale free avalanches \cite{LMW10,MW15} whose statistical properties can be studied in detail. 

The purpose of the present work is to develop an algorithm allowing to explore adiabatically the isostatic landscape of the perceptron problem with linear cost function. We define an athermal adiabatic  procedure, similar in its architecture to the one developed to follow packings of hard spheres under strain \cite{Ro00, CR01, Roux00, CR02, CR00, LDW13simu}.  The evolution of the system is tracked by computing the points where marginally satisfied constraints (contacts) destabilize. The avalanche that is triggered is followed till a new stable configuration is found. 

We use this algorithm to start from a configuration at jamming and perform a progressive compression to enter the jammed phase. 
We analyze both the situation in which the compression is performed from a convex (meaning not-critical) jamming point as well as when the system is prepared at the non-convex jamming point.  The two situations are rather different: while in the former case the jamming point can be followed with continuity when it is progressively compressed, in the latter such \emph{state following} situation is not possible since local minima are marginally stable and undergo avalanches.  The evolution of the energy landscape is rather chaotic with crossing between different minima at different energy levels. 
We characterize the statistics of avalanches showing that we get the same as for strained packings of hard spheres.  Finally the algorithm allows to investigate the properties of the unjamming transition.  Remarkably, the unjamming transition in this case does not fall in the general framework developed 
in \cite{OSLN03,WSNW05,GLS16} where the linear cost function we are using corresponds to a marginal situation.  Thermodynamic quantities develop logarithmic singularities at jamming and new considerations are necessary to understand this behavior. 

\section{The spherical perceptron with linear cost function}
The perceptron is one of the oldest models in machine learning \cite{R58} to perform binary classification of patterns. 
The problem of classification of random patterns has been 
studied extensively in statistical mechanics starting from the '80s \cite{Ga87, Ga88,DG88}.
Our main interest here is that it provides the simplest model of random constraint satisfaction problem (CSP) with continuous variables \cite{FP16}. While CSPs with discrete variables are important to model combinatorial optimization problems which are central in algorithmic complexity theory, problems with continuous variables are less studied from this point of view, but are central from the optimization viewpoint. The perceptron problem that we consider here has been recently investigated for the same reasons in the the computer-science/mathematical literature, see \cite{ST03, Sto13, ES20}.
For our purposes, we define the perceptron problem  in terms of an $N$-dimensional state vector $\x$ on the $N$-dimensional sphere $|\x|^2=N$ subject to $M=\alpha N$ random soft constraints.  The constraints, are build from a set of $M$ $N$-dimensional vectors $\pat_{\mu}\in \R^N$ with $\mu=1,...,M$, usually called \emph{patterns} that we take as random points on the sphere. 
Each component $\xi_\mu^i$, $i=1,\ldots N$ is a Gaussian random number with zero mean and unit variance.
The total number of random vectors, $M$, scales with the dimension of the phase space as $M/N=\alpha$ and $\alpha$ is a ${\cal O}(1)$ control parameter of the problem.  Given the vector $\x$ and the set of patterns and a real number $\sigma$, one can construct $M$ gap variables $h_{\mu} = {\pat_\mu \cdot \x}/\sqrt{N}-\sigma$ and a cost function, or Hamiltonian, as
\begin{equation}
H[\x] = \sum_{\mu} |h_\mu|^a \theta(-h_{\mu})
\label{hamiltonian}
\end{equation}
where $a\ge 0$ is a positive exponent. The gap variables represent soft constraints as according 
to Eq.~\eqref{hamiltonian}, there is an energy penalty whenever a gap is negative. 
The variable $\sigma$, that quantifies the difficulty in satisfying the constraints, is an important control parameter in the problem and is called
margin\footnote{Also this name comes from the machine learning interpretation. In that context one uses the variable $\sigma$ to make sure that the hyperplane separating the points is at least $\sigma$ far from each point.}. In the classification problem, $\sigma>0$ and each constraint defines a convex zero energy region around a pattern. For $\sigma<0$
the problem is still well defined, but it cannot be interpreted as a classification problem any more. Rather
 the patterns can be seen as obstacles that the configuration of the system should avoid. The zero energy region of each pattern is non-convex, 
$|\sigma|$ is the radius of exclusion around each obstacle and can be seen as a density parameter analogous to packing fraction in the packing of spheres (see \cite{FPUZ15, FPSUZ17} for a detailed comparison between the sphere model and perceptron). 

Regarding the exponent $a$ in the cost function, in this paper we concentrate on the case $a=1$, 
and the amount of energy that is payed is linear in the absolute value of the gap. This linear case separates the case where 
$H$ is convex in each of the $h$ from the case it is concave. Correspondingly, the Hamiltonian is not differentiable in $h_\mu=0$, and contacts can sustain forces without any energy increase. We have discussed these properties and some of the consequences in detail both for the perceptron and for soft spheres in Refs. \cite{FSU19,FSU20}.  Here follows a summary of the main results of the analysis.

\begin{figure}
\centering
\includegraphics[scale=1.0]{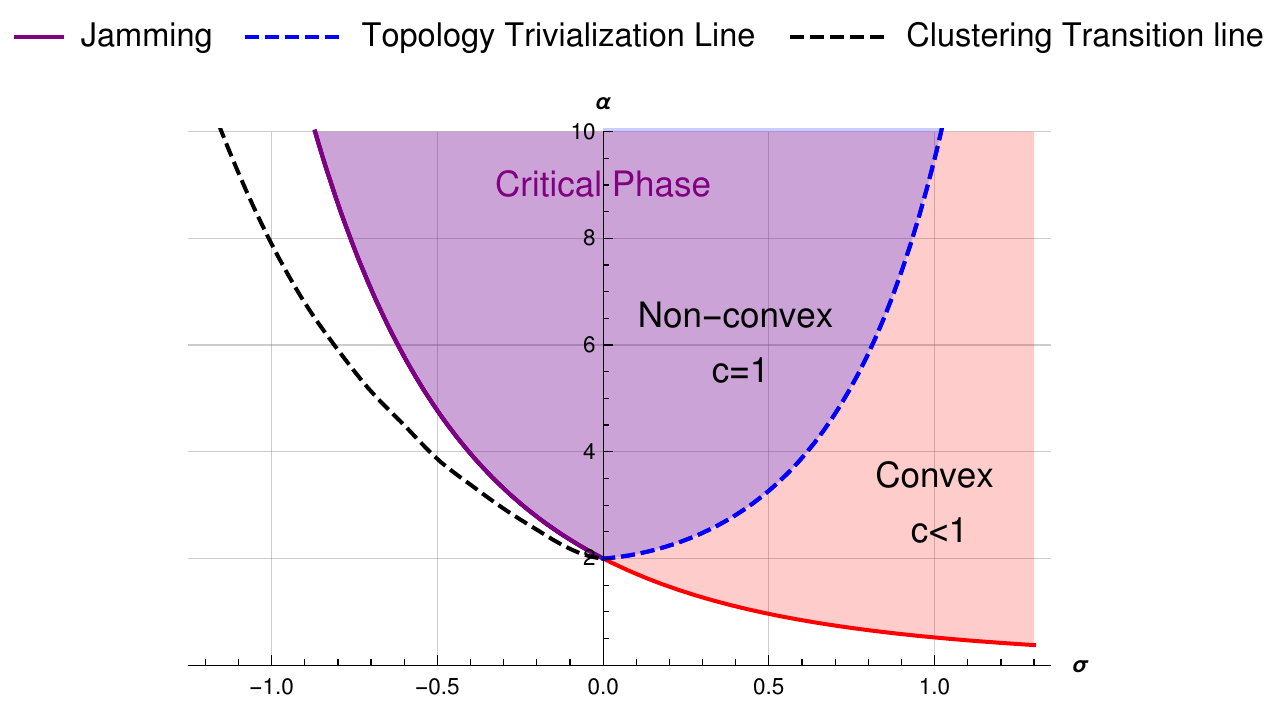}
\caption{The phase diagram of the spherical perceptron problem with linear cost function \cite{FSU19}. The blue dashed curve is a topology trivialization transition line (that coincides with the onset of Replica Symmetry Breaking). Above this line the landscape is non-covex with many local minima while below it is convex with just one unique minimum. The isostaticity index, defined as the number of contacts divided by $N$ is $c=1$ in the glassy/non-convex phase while $c<1$ in the convex phase. In the glassy phase one has isostatic minima which are marginally stable. In the SAT phase, we indicated with a black dashed line the point where the solutions to the satisfiability problem gets clustered in far away lumps and replica symmetry breakes.
}\label{phase_diagram}
\end{figure}

The perceptron model is exactly solvable, e.g. with the replica method,  its ground state of $H$ has been studied in \cite{GG91,MEZ93,FP16,FPSUZ17,FSU19}.  
In Fig.\ref{phase_diagram} we reproduce its phase diagram in the hinge loss case.  

At small $\alpha$ and $\sigma$ the problem is Satisfiable (SAT); configurations exits where all gaps are positive and there is a degeneracy of ground states of $H[\x]$ with zero energy. Increasing $\alpha$ or $\sigma$, the ground state meets a jamming transition, beyond which 
the problem becomes Unsatisfiable (UNSAT),  where some gaps are negative in the ground state and the energy density is positive. The properties of the SAT phase, and the jamming line itself are independent of the choice of the loss function. We focus here on the nature of the UNSAT or jammed phase,  which, conversely, is deeply affected by the choice of the exponent $a$. 

For $a\ge 1$ the UNSAT phase is divided in a region where landscape is effectively convex and the cost function (\ref{hamiltonian}) has a unique minimum, and of a region which is non-convex and glassy with marginally stable minima \cite{FSU19}. Technically the two UNSAT phases, with rather different properties, are separated by a replica-symmetry-breaking (RSB) transition line.  Within this framework, the case of hinge loss emerges as a particularly interesting case. This can be realized looking at the distribution of the gap variables. As soon as one enters the UNSAT phase, for $a>1$ the probability of exact contacts $h_c=0$ is zero. Contacts are destabilized by any small applied forces and one finds either positive gaps that do not contribute to the energy or negative gaps.  For $a=1$ conversely, contacts $c$ can sustain forces $f_c$ provided that they lie in the interval $f_c\in(0,1)$. The jammed configurations are then characterized by a finite density of gaps that are identically zero. In the convex phase the ground state of the problem is {\it hypostatic}: the total number $C$ of such contacts is sufficient to insure mechanical stability and $C/N<1$. In the non-convex phase conversely,  minima are {\it isostatic} meaning that  $C=N$.

Analogously to what happens on the non-convex jamming line, isostaticity implies marginal stability.
The properties of  isostatic configurations are characterized by a critical distribution of the gaps, $\rho_h(h) = \frac 1 M\sum_{\mu} \delta(h-h_\mu)/M$, which for both small positive and negative value presents a critical power form $\rho_h(h)= A_\pm |h|^{-\gamma}$ where
with $A_{\pm}$ two constants and $\gamma$ is a parameter independent exponent that coincides critical exponent that controls the positive gap at jamming $\gamma= 0.41269\ldots $ \cite{CKPUZ14,CKPUZ13}.  Therefore local minima are such that there is an isostatic number of contacts, namely gaps that are strictly equal to zero, and an abundance of both small positive and negative gaps. Correspondingly, the contact forces $f_\mu\in (0,1)$, whose empirical distribution $\rho_f(f) = \sum_{\mu}\delta (f-f_\mu)/C$ has support in $f \in (0,1)$ and in the thermodynamic limit develops two pseudogaps close to the edges of the support $\rho_f(f) \sim B_0 f^\theta$ for $f\to 0$ and $\rho_f(f) \sim B_1 (1-f)^\theta$ for $f\to 1$.
Again, the amplitudes $B_0$ and $B_1$ are two constants  while numerically we find that $\theta$ is close to $\theta=0.42311\ldots$ which can be obtained by assuming a continuous RSB solution. The exponent $\theta$ does not depend on the parameters and is equal to the critical exponent controlling the small contact forces at the jamming point \cite{CKPUZ14,CKPUZ13}. Isostaticity 
makes local minima of Eq.~\eqref{hamiltonian} marginally stable in the thermodynamic limit, and barely stable for finite sizes. In order to understand qualitatively why this is so, we can consider the system sitting in one of such isostatic minima. Isostaticity and non-convexity imply that if one of the contacts is removed, the system looses mechanical stability and moves away.  Due to the pseudogaps in the force distribution, a perturbation that is vanishingly small in the thermodynamic limit is sufficient to push one of the contact forces outside the stability interval $(0,1)$. Force balance is now off by one contact and the system moves away. A new equilibrium configuration is obtained at the expense of rearrangements.  New contacts need to be formed either by  positive gaps that become contacts, or by negative gaps that cross zero. The probability of having a system off by a contact when perturbed is controlled by the two pseudogaps in the force PDF close to zero and one. Analogously the distance the system travels from the unstable configuration to a new stable one is controlled by the abundance of the small positive and negative gaps, which are most likely to form new contacts. The two effects counterbalance and, as we have shown in \cite{FSU19, FSU20}, the system is marginally stable. This argument is similar to the one used to rationalize the response to shear perturbations in jammed hard spheres \cite{CR02,DDLW15}, which also lead isostatic configurations into isostatic configurations and, as we will see, can be analyzed qualitatively in a similar way. 

In the following we would like to model quasi-static compression-decompression dynamics within the jammed phase, follow isostatic local minima till the verge of stability and then, when  the minimum is lost, understand the plastic events that lead to the next one. 
Compression could be achieved modifying $\sigma$ or increasing the number of constraints.
%\footnote{A third possibility would be to add a degree of freedom to the system $N\to N+1$.}.  
However, any increase of $\sigma$, however small, could not be considered a small perturbation, as it would break all the contacts at the same time and imply a complete rearrangement of the system.
%\footnote{We will see that the same is true for adding a degree of freedom.}. 
In the same way the addition of constraints would modify the landscape in a discontinuous way. It is then convenient to define a pressure variable $p$, Legendre conjugate to $\sigma$ in the energy, and consider $\sigma$ as a dynamic variable on the same level as the $\x$. We have then to minimize the Legendre function $L[\x,\sigma]=H-N\sigma p$.
 As we will see, small enough increases in pressure do not destablize the minima.

\subsection*{Constitutive equations for local minima, a Lagrangian formulation}

In order to write the conditions of minimum that properly take into account contacts, it is useful to define a Lagrangian function as a function 
of the system's variable and the forces, thought as Lagrange multipliers that enforce the contacts
\begin{align}
    \fe = 
    \sum_{o\in \Ov} \left(\sigma - \frac{\pat_o \cdot \x}{\sqrt{N}}\right) +
    \sum_{c\in \Co} f_{c} \left(\sigma - \frac{\pat_c \cdot \x}{\sqrt{N}}\right) + 
    \frac{\lm}{2} \left(|\x|^2 - N\right) 
    - p \sigma N
    \label{lagra}
\end{align}
where we have defined the set of overlaps and contacts, respectively
\begin{equation}
\Ov = \{o : h_o<0\} \ \ \ \ \Co = \{c : h_c=0\}.
\end{equation}
We often indicate contacts with indexes $c$ and overlaps with indexes $o$ leaving understood the set they belong to.  
The first term of Eq.~(\ref{lagra}) is just the Hamiltonian and by itself it pushes $\sigma$ to be small in order to make the gaps positive. The second term instead is made of the Lagrange multipliers $f_c$ associated to the contacts. The third term is another Lagrange term which enforces the spherical constraint on $\x$ through the Lagrange parameter $\lm$. Finally, the fourth term, for $p\geq 0$, fixes $\sigma$ by compressing the system. The tradeoff between the energetic payoff and the last term is set by the pressure $p$. 
Once fixed the sets $\Ov$ and $\Co$, the extrema of the Lagrangian satisfy the first-order conditions
\begin{equation}
\left\{\begin{split}  
\partial_{X_i} \fe &= 
    \sum_{o} \frac{-\xi_{o,i}}{\sqrt{N}} + 
    \sum_{c} f_{c} \frac{-\xi_{c,i}}{\sqrt{N}}+
    \lm X_i=0\\
    \partial_{\sigma} \fe &= 
    \sum_{o} 1 + 
    \sum_{c} f_{c} -
    p N=0
\end{split}
\right.
\;\;\;
\left\{
\begin{split}
\partial_{f_c} \fe& = -h_c=\sigma - \frac{\pat_c \cdot \x}{\sqrt{N}}=0 \;\;\;\forall c\in \Co\\
 \partial_{\mu} \fe& =\frac 1 2 (\x^2-N)=0
\end{split}
\right.
\label{grad_zero}
\end{equation}
The first set of these equations states the force balance conditions on each of the variables and fact that the pressure is the average force due to 
contacts and overlaps. The second sets of equations describes the conditions that $h_c=0$ for all contacts and the spherical constraint for $\x$. 
From the costitutive equations (\ref{grad_zero}) we can derive a '1st principle'-like relation between $p$, $\sigma$, $\mu$, and the intensive energy due to the overlaps  $e={\sum_{o} |h_{o}|}/{N}$ . Indeed, if we consider  Eqs.~\eqref{grad_zero} and we multiply them by $\x$ we obtain
\begin{align}
\label{pe}
0=\sum_{i} X_i \partial_{X_i} \fe = N (\mu -p\sigma +e)
\end{align}
and taking the derivatives we get
\begin{align}
    p= \frac{\de e}{\de \sigma}\bigg|_{\p,\lm} \ \ \ \ \ \ \sigma = \frac{\de\lm}{\de\p}\bigg|_{\sigma,e}\:.
\end{align}
The first relation is consistent with the definition of pressure. The second one tells that $\lm$ has a minimum for $\sigma=0$: indeed we know that $\lm =0$ at jamming and $\lm <0$ for $\sigma<0$ \cite{FSU19}.  It so happens that $\lm>0$ in the convex phase and $\lm<0$ in the non-convex phase \cite{FSU19}, therefore we have $E<p\sigma$ in the convex phase and $E>p\sigma$ in the non-convex phase. According to (\ref{grad_zero}), the pressure cannot exceed the value of $\alpha$, corresponding to having all negative gaps; in correspondence to that value, $\sigma$ should be divergent. 

Note that the total number of equations is $N+C+2$, which coincides, as it should, with the number of variables. Generically, for any 
given disjoint sets $\Ov$ and $\Co$ one should expect one or more solutions to these equations. However, only the points where
$h_o<0$ for all $o\in \Ov$ and $f_c\in (0,1)$ for all $c\in \Co$ correspond to physically stable solutions. For a given sets of external parameters, only 
appropriate choices of the sets $\Ov$ and $\Co$ give rise to physical solutions. We need therefore an algorithm that allows to update these sets and find physical solutions as the pressure is changed. 

In order to proceed, it is useful to define a more compact notation grouping variables and Lagrange multipliers. 
We renumber the patterns in such a way that the set of contacts becomes $\Co=\{1,...,C\}$
and define a $C+1$ dimensional vector  $\fv = \begin{bmatrix} f_1 \\ ... \\ f_C \\\mu\end{bmatrix}$ so that the first 
$N+1$ extrema equations can be formally written under the form of linear conditions for $\fv$,
\begin{eqnarray}
  \label{eq:1}
    \nabla \fe = \begin{bmatrix}\partial_{X_1}\fe  \\ ... \\ \partial_{X_N}\fe   \\\partial_{\sigma}\fe  \end{bmatrix}
\equiv \Smat \fv - {\bf v}=0. 
\end{eqnarray}
Where the $(N+1)\times (C+1)$ matrix $\Smat$ is defined as 
\begin{align}
    \Smat =
    \begin{bmatrix}
    \frac{-\pat_1}{\sqrt{N}} & \frac{-\pat_2}{\sqrt{N}} & ... & \frac{-\pat_C}{\sqrt{N}} & \x \\
    1 & 1 & ... & 1 & 0
    \end{bmatrix}\:.
\end{align}
and the $N+1$ dimensional vector
  \begin{eqnarray}
    \label{eq:3b}
   {\bf v}=\begin{bmatrix}\frac 1 {\sqrt{N}}\sum_o \pat_o  \\ pN  -O \end{bmatrix}
  \end{eqnarray}
 Notice that for choices of $\Co$ such that $C=N$, we formally seek an isostatic configuration. One can first solve the second set of equations for $\x$ and $\sigma$ and then, for fixed values of these variables find the corresponding forces from the first set of equations. 

If this configuration is also a physically meaningful one, then $f_c\in (0,1)$ for every $c\in \Co$ and all gaps in $\Ov$ are negative. In the following section we describe an algorithm to find new physical states after the destabilization of a physical isostatic configuration.

\section{Algorithm: surfing on isostatic minima} 
We will mainly focus on studying the quasi-static compression of a configuration starting from an isostatic jamming point and entering in the non-convex jammed phase.  Therefore, according to the phase diagram, we will fix $\alpha>2$ and start a compression from the jamming line.  Note that the location of the jamming point is algorithm dependent, as in the non-convex RSB phase there is no guarantee to find the absolute minimum of the system.  We want to design an algorithm that given an isostatic configuration at jamming is able to follow the same configuration when the pressure is progressively increased.  The step zero of the algorithm is therefore to produce a configuration at jamming. Given that we will describe how to perform a compression.  
\subsection{Step zero: producing a configuration at jamming} 
An easy way to produce a configuration at jamming is to follow \cite{FSU19} and to consider the smoothed version of the Lagrangian \begin{align}
    \fe_{\epsilon} = 
    \sum_{o} \left(\sigma - \frac{\pat_o \cdot \x}{\sqrt{N}}\right) +
    \sum_{c} \frac{1}{2\epsilon} \left(\sigma - \frac{\pat_c \cdot \x}{\sqrt{N}}\right)^2 + 
    \frac{\lambda}{4} \left(|\x|^2 - N\right)^2 
    - p\sigma N
    \label{eq:smoothed}
\end{align}
where we have introduced a regularization of the interaction at contacts and of the spherical constraint. The extrema of the Lagrangian Eq.~\eqref{lagra} are recovered from the one of the smoothed one in an appropriate limit $\epsilon\rightarrow 0$ and $\lambda\rightarrow \infty$.  Once the number of patterns $\alpha$ and a value of $p$ \emph{small enough} are chosen, we run a gradient descent on the cost function in Eq.~\eqref{eq:smoothed} with the degrees of freedom $\xs$: this produces a configuration $\xs_J$. 
The pseudocode corresponding to this part of the algorithm is in Fig.\ref{alg:jamming}.
\begin{figure}
\begin{algorithm}[H]
	\SetAlgoLined
	Initialize $(\x,\sigma)$ at a random with the contraint $|\x|^2=N$\;
	set $p$ small enough: $p<p_J(N)$\;
	initialize $\epsilon=\epsilon_{i}$\;
	\While{$\epsilon>\epsilon_{f}$}{
	$(\x,\sigma) \gets \arg\min \fe_{\epsilon}\left(\x,\sigma\right)$\;
	$\epsilon \gets \epsilon/2$\; 
	}
	\KwResult{$(\x,\sigma)$ is a configuration at the jamming point.}
	\caption{Step zero: producing a configuration at the jamming point}
	\end{algorithm}
	\caption{The pseudocode to produce random configurations at jamming. The minimization of $\fe_{\epsilon}\left(\x,\sigma\right)$ is performed with an approximated conjugate-gradient method (L-BFGS) \cite{lbfgs-BN95,SciPy}. The parameters we used for the system sizes we have studied are: $\epsilon_{i}=10^{-2}$; $\epsilon_{f}=10^{-8}$; $p=0.1$.}
	\label{alg:jamming}
\end{figure}

\begin{figure}
\centering
\includegraphics[scale=0.7]{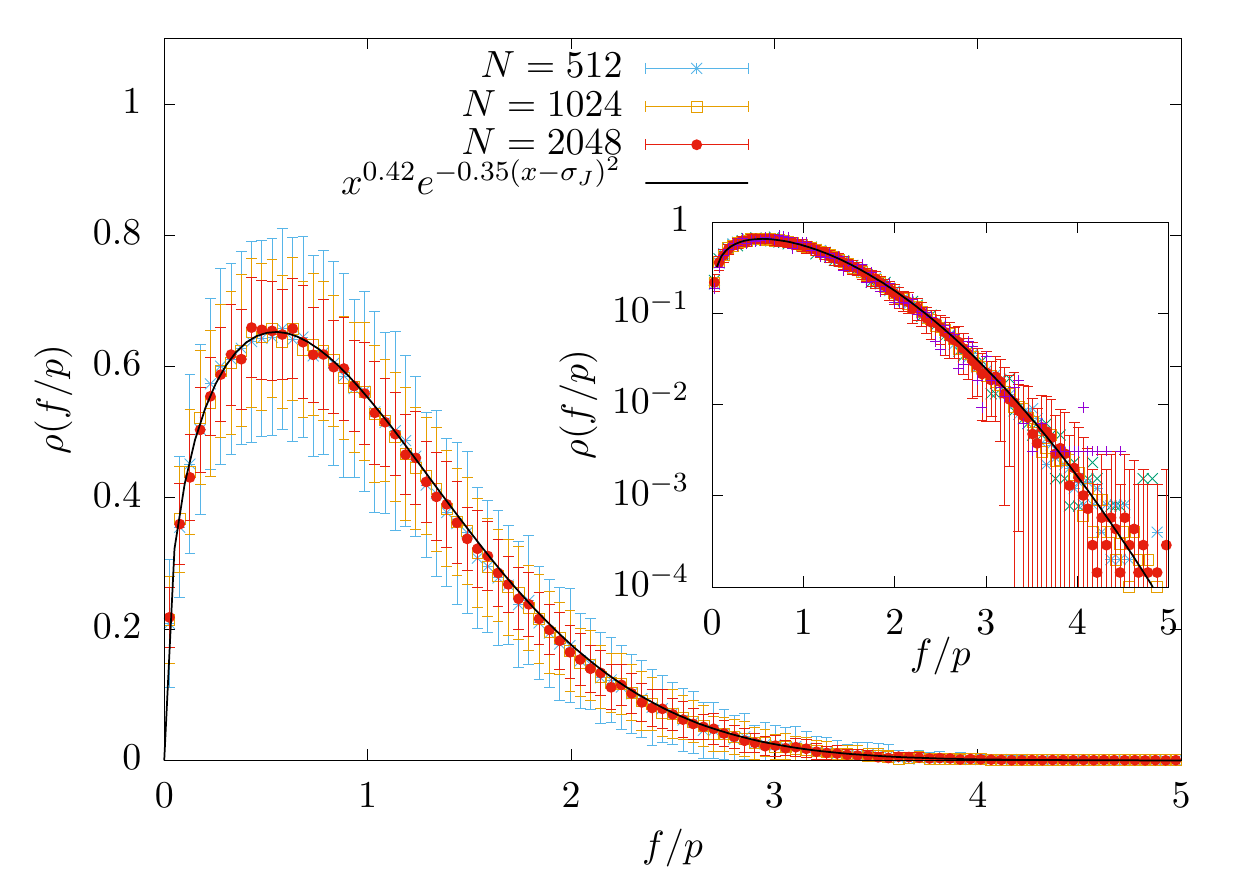}
\includegraphics[scale=0.7]{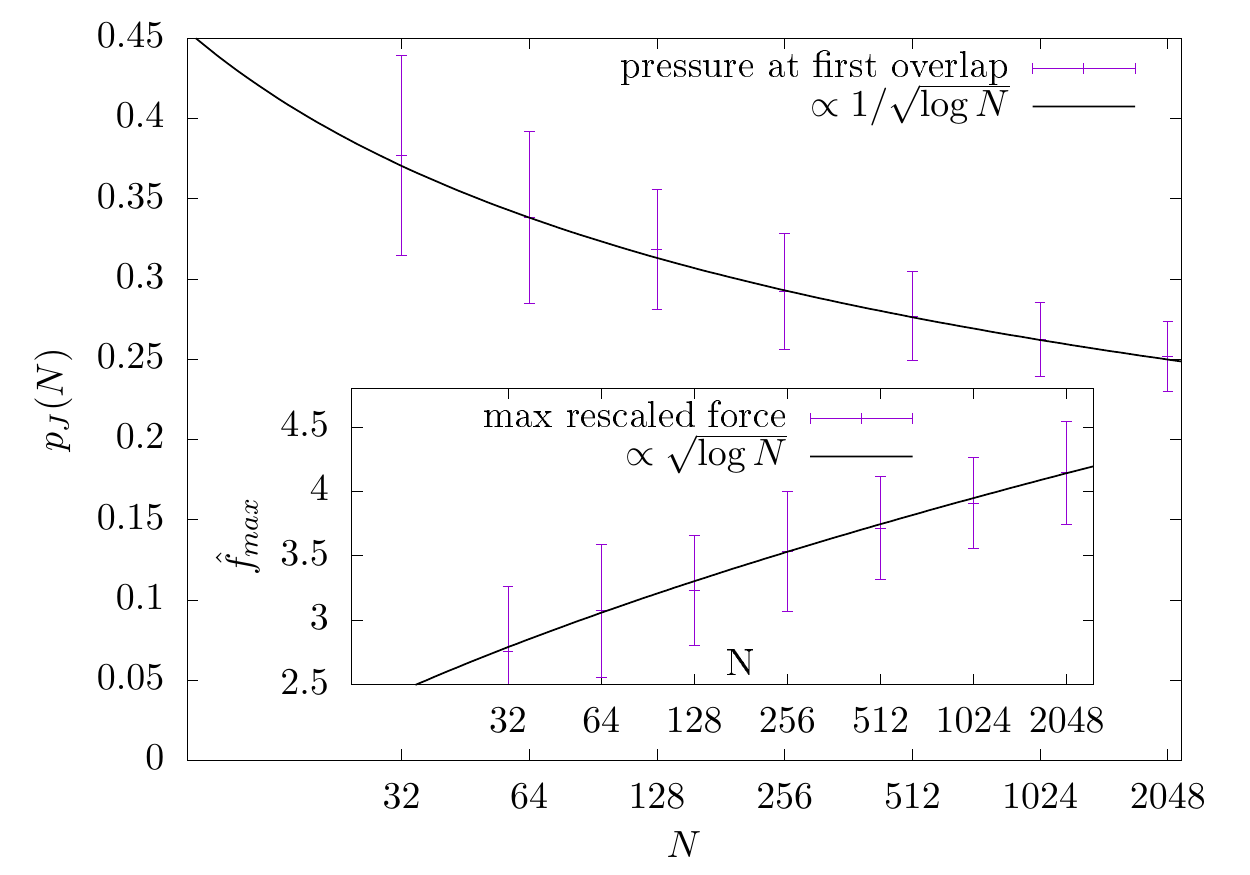}
\caption{Left panel: The scaled force distribution at jamming for different system sizes at $\alpha=4$. Inset: zoom on its tail. In both cases we plot with a black line a simple scaling form that retains the critical properties fo the distribution for $\hat f\to 0$ as well the exponential behavior for large forces. Right panel: the scaling of $p_J(N)$ as a function of the system size. In the inset we plot the maximum scaled force $\hat f_{\rm max}$ as a function of the system size. In black lines we show a logarithmic fit. Only at jamming we also include the results of simulations for $N=2048$.}
\label{fig:force_jamming}
\end{figure}

  In the thermodynamic limit, jamming points are only stable for pressure $p\to 0$. However, as it will become clear in a moment, for a finite size system configurations at jamming can sustain small pressures without moving, and they can be compressed till the point where the largest of the contact forces leaves the stability interval $(0,1)$. Therefore a finite size configuration at jamming is stable in an interval of pressure values $p\in [0,p_J(N)]$ where $p_J(N)\to 0$ as $N\to \infty$. Given a configuration at jamming, we can compute the contact forces from Eq.~\eqref{grad_zero}: For $p$ strictly equal to zero all the contact forces are zero.  Increasing $p$, since the set of overlaps is empty $\Ov=\emptyset$, the solution of the linear system has the form $f_c=p \hat{f}_c$ with $\hat{f}_c$ independent of pressure.  The force distribution therefore progressively invades the stability interval $(0,1)$, and the solution is stable till pressures $p_J(N)=1/\hat{f}_{\rm max}$ where the largest contact force exits the stability interval and we enter the jammed phase. In Fig.~\ref{fig:force_jamming}-left panel, we show the empirical distribution $\rho_f(\hat{f})$ of the rescaled forces $\hat f=f/p$ at jamming for $\alpha=4$.

The distribution displays the critical pseudogap for $\hat f\to 0$  {\cite{FPSUZ17} in analogy with hard spheres at jamming, see \cite{CKPUZ17} for a review.} In addition, similarly to what found in \cite{OSLN03} for jamming points in spheres,  we empirically observe a large argument tail which is compatible with a Gaussian. 
Usual extreme value statistic arguments, supposing independence of the contact forces, imply that $\hat f_{\rm max} \sim \sqrt{\ln N}$, or  $p_J(N)=1/\hat{f}_{\rm max}\sim (\ln N)^{-1/2}$. Fig.\ref{fig:force_jamming}-right panel shows that such a  scaling is in agreement with numerical simulations, and gives rise to 
rather large critical pressures for the system sizes we have studied. When the pressure reaches $p=p_J(N)$ the system becomes unstable because at least one of the contact forces goes out from the stability interval $(0,1)$ and we enter the jammed phase.

Notice that the argument for finite volume stability of isostatic configuration under small pressure changes extends without changes into the jammed phase. In fact, also for isostatic jammed configurations, despite now the forces are not simply proportional to pressure, we still have that for fixed $\xs$ a small pressure change $\delta p$ induces small force changes $\delta f_c\sim \delta p$. 
{Since generically all forces are strictly smaller then one, this does not cause destabilization if $\delta p$ is small enough. }

\subsection{Inside an avalanche: simulating a plastic event}
Once the pressure destabilizes the configuration at jamming we enter the UNSAT phase. Let us describe the algorithm in the case of compression, the case of decompression would be specular.  Suppose that we are in an isostatic configuration at jamming or in the jammed phase. We can increase the pressure
$p\to p+\delta p$,  until the larger of the contact forces computed from Eq.~\eqref{grad_zero} hits the right edge of stability support $(0,1)$. The corresponding contact, corresponding say to pattern $\pat_k$, is not stable anymore and becomes an overlap. This destabilization of the weakest contact leads to an avalanche. The system then needs to moves away through a plastic event, where the pattern of contacts changes. Following the destabilization, the vector $\xs$ is modified in the direction of the unstable mode {\cite{CR02}}: the unique direction that preserves the $N-1$ remaining contacts. The motion can be followed till a new contact is formed; we need at this point to update $\Co$, recompute the forces and stop if all the forces are in the physical support. It may  happen however that one or more of the forces get unphysical values. We should then iterate the procedure of breaking the unphysical contacts and the avalanche must go on. If the number of contacts $C<N-1$, the 'soft-mode manifold' of moves that conserves the physical unbroken contacts, is multidimensional and we should decide in which direction to move. One possibility is simply to use the present values of the forces and follow the projection of the gradient on the soft manifold. We found however that this procedure converges very slowly. A better prescription can be obtained observing that, while for $C<N$ in general the gradient $\nabla \fe$ cannot be made equal to zero by a choice of the forces, it can be rendered orthogonal to the 'hard manifolds' of moves that modify the contacts. This is achieved imposing that 
\begin{eqnarray}
  \label{eq:5}
  S^T\nabla \fe =S^T S \fv-S^T {\bf v}:
\end{eqnarray}
which is a system of $C+1$ equations that can be solved for $\fv$, the vector of forces and $\mu$. Let us call  $\fv^*$ the solution and $\nabla^*\fe$ the corresponding gradient.
It is natural then to define moves of $\xs$ that follow  $\nabla^*\fe$ until a new contact is created. 
\begin{eqnarray}
 \delta \xs=-\eta\nabla^*\fe
\end{eqnarray}
where $\eta$ is chosen in such a way that a new contact appears, either a gap that closes, or an overlap that becomes a contact. We remark that 
the choice (\ref{eq:5}) corresponds to minimize $(\nabla \fe)^2$ with respect to $\fv$: the resulting variation $\delta\xs$
is the smoothest one in the soft manifold. 
We have now a new configuration $\xs\to 
{\begin{bmatrix} \x+\delta \x\\ \sigma+\delta\sigma \end{bmatrix}}$,
but the spherical constraint is not respected by ${\bf X}+\delta {\bf X}$. A configuration 
on the sphere can be simply obtained by contemporary rescaling of ${\bf X}$ and $\sigma$ which by definition does not affect the contact conditions. 
This procedure of adding and removing contacts can be iterated until a new physical isostatic configuration is found.  We are then ready 
for further pressure increase. 
The pseudocode to perform such compression steps is in Fig.\ref{alg:compression}.
\begin{figure}
\begin{algorithm}[H]
	\SetAlgoLined
	Consider a stable configuration at $(\x,\sigma)$ and $p_0$\;
	$\delta p^* \gets$ compute critical pressure variation at \underline{fixed} $(\x,\sigma)$\; 
	$p \gets p_0 + \delta p^* + \delta\delta p$\;
	$\fv^* = \underset{\fv,f_c\in[0,1]}{\arg\min}| \nabla\fe\left(\fv\right)|^2$ at fixed configuration $(\x,\sigma)$\;
	$\nabla^*\fe\gets \nabla \fe\left(\fv^*\right)$ \;
	\While{$|\nabla^*\fe|>\hat \tau$}
	{	
		$\delta h_{\mu} = -(\pat_{\mu}/\sqrt{N},-1) \cdot \nabla^*\fe$\;
		$t^* \gets \underset{\mu}{\min} \{t_{\mu}\ :\ h_{\mu} + t_{\mu}\delta h_{\mu} = 0 \land t_{\mu}>0 \land \mu\notin\Co\}$\;
		$\eta \gets t^*$\;
		\If{$\lm>0$}{
			$\qquad \eta \gets \min\{t^*,\frac{1}{\lm}\}$\;
			} 
		$(\delta\x,\delta\sigma) \gets -\eta \nabla^*\fe$ \;
		$K\gets \sqrt{N} / |\x+\delta\x|$\;
		$ (\x, \sigma) \gets K\left[(\x, \sigma) + (\delta\x,\delta\sigma)\right]$\;
		get the new sets $\Co$, $\Ov$\;
		$\fv^* = \underset{\fv,f_c\in[0,1]}{\arg\min}| \nabla\fe\left(\fv\right)|^2$ at fixed configuration $(\x,\sigma)$\;
		$\nabla^*\fe\gets \nabla \fe\left(\fv^*\right)$ \;
	}
	\KwResult{$(\x,\sigma)$ is a stable configuration at pressure $p_0 + \delta p^* + \delta\delta p$}
	\caption{Compression step}
\end{algorithm}
\caption{The critical pressure variation $\delta p^*$ is computed as the positive $\delta p$ so that $\delta{\bf f}^TS^T S =(\bm{0}, \delta p)S^T$  gives $f_c+\delta f_c = 1$ or $f_c+\delta f_c = 0$ for only one contact force $c \in \Co$ (this coincides with the point in which the first one among the perturbed forces gets out of the stability interval). Setting the new pressure at $p_0 + \delta p^*$ would put the configuration at the verge of instability but still with zero gradient. Therefore we add a small destabilizing push $\delta\delta p$ that we fix to $10^{-8}$. Computing $\fv^* = \underset{\fv,f_c\in[0,1]}{\arg\min}| \nabla\fe\left(\fv\right)|^2$ is a constrained least-squares problem that we solve using the algorithm described in \cite{lawson1995lsq} and implemented in \cite{SciPy}. The tolerance $\hat \tau$ for the gradient is set to $10^{-11}\sqrt{N}$.}
	\label{alg:compression}
\end{figure}

\subsubsection*{The algorithm in the convex phase}
In the convex phase, since the landscape has a unique minimum, any minimization algorithm would produce the same configuration upon compression. 
For example, one could use  gradient descent minimization at each time the pressure is changed.  While this is possible, it remains interesting  to study the adiabatic contact-breaking dynamics in this case.   

Differently from the isostatic case, when following the minimum for fixed contact and overlap sets  $\Co$ and $\Ov$, in the hypostatic case 
the contact conditions only specify $C<N$ variables and both the contact forces and the position move as a consequence of the pressure variation. 
As a consequence, in principle, the system can become unstable in two ways, either by a contact force that exits its support, or by a gap that changes 
sign. As in the isostatic case, we should follow a minimum for fixed $\Co$ and $\Ov$ to the first instability and then iteratively change the contact and overlap set till a new stable solution is found. 

Suppose to start from the minimum at pressure $p$ and increment the pressure to $p+\delta p$. We can find the variation of the force vector projecting the equations on the patterns as in the previous section:
\begin{eqnarray}
  \label{eq:25}
  S^T S \delta {\bf f} =S^T \begin{bmatrix} \bm{0} \\ \delta p \end{bmatrix}.
\end{eqnarray}
As in the previous section we can determine the minimal pressure variation that leads one of the forces outside the physical support. 
Once solved the (\ref{eq:25}), the gradient $\nabla \fe$ is orthogonal to the patterns and we can determine $\delta X$ by the condition 
\begin{eqnarray}
  \label{eq:26}
  \delta X_i=-\eta \nabla_i \fe[{\bf f}+\delta \fv,{\bf X},\sigma]\:.
\end{eqnarray}
Notice that if we choose $\eta=1/\mu$, the gradient is zero in the new configuration. If this gives rise to a physical solution we can make this choice. If instead we find that there at least one gap that has changed sign, we reduce $\eta$ to the value where till no gap has changed sign. 
The final step, consists in rescaling the variables together with $\sigma$ to bring the new configuration on the sphere. We need now to change the set $\Co$ and $\Ov$ and iterate the procedure till a stable configuration is found.

\section{Simulating compression}

\begin{figure}
\includegraphics[scale=0.7]{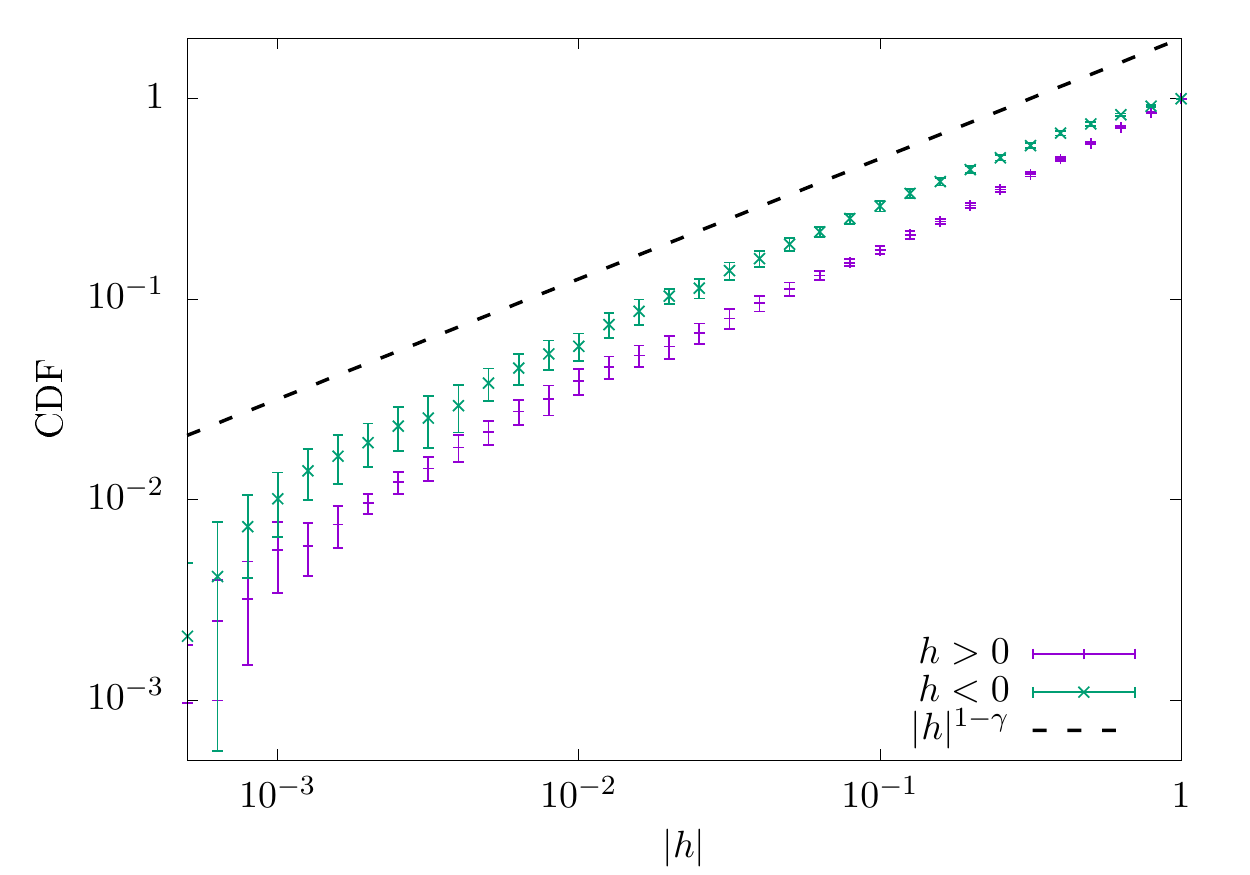}
\includegraphics[scale=0.7]{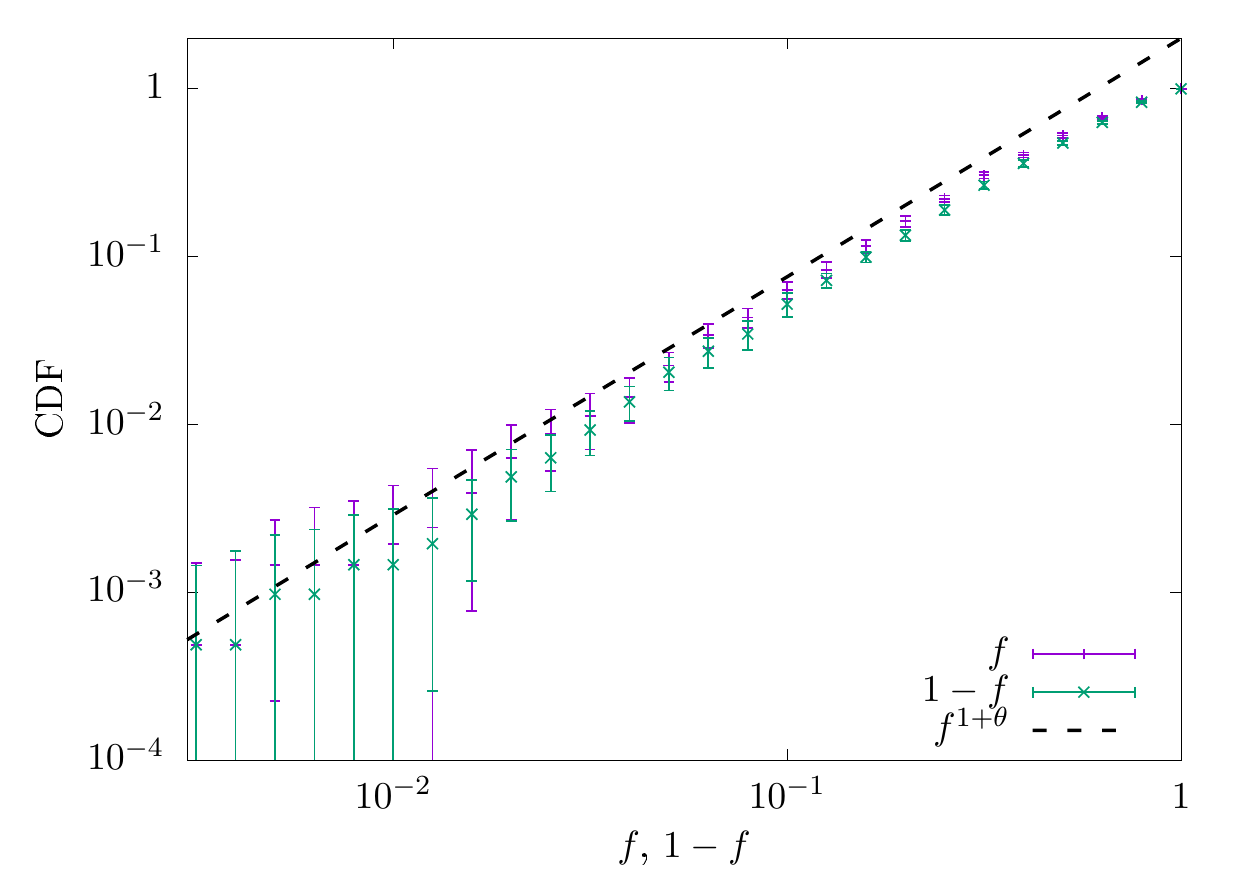}
\caption{\emph{Right panel}: the cumulative distribution of forces close to the two edges of the stability interval, namely $f=0,1$. \emph{Left panel}: the cumulative distribution of both positive and negative gaps close to the origin. Both forces and gaps display singular behavior of the same kind of the local minima found in \cite{FSU19}. The figure is produced out of 10 samples, with $N=1024$ at $\sigma=0$ and $p=1.37$.}
\label{forces_gaps}
\end{figure}

In this section we discuss the results obtained by running the algorithm we just presented. 
We simulated compression from jamming point for the perceptron for values of $\alpha$ as specified below. 
We mainly simulated systems with $\alpha>2$ where the jamming point lies in the non-convex domain of the phase diagram, but for comparison also values $\alpha<2$ were considered. 
In the non-convex region we first show that local minima that are found by the adiabatic compression algorithm display the same universal properties as the ones found using gradient descent-like minimization \cite{FSU19}. Indeed local minima are found to be isostatic and in  Fig.\ref{forces_gaps} we show the singular distribution of forces and gaps in the jammed phase at finite pressure. This singular behavior is both consistent with mean field theory as well as with the numerical simulations of \cite{FSU19}.
Therefore our surfing algorithm explores minima that share the same universal features of local minima obtained by different greedy local algorithms.

We simulated sizes $N=64, 128, 256, 512, 1024$, all quantities were averaged over 100 independent samples. 

We will first show the behavior of bulk physical quantities across the full compression cycle
and then we will consider the statistics of avalanches. Finally we will discuss what happens close to unjamming.

\subsection{Following jammed configuration across the phase diagram: the topology trivialization transition}
We are interested in studying the behavior of the system following a compression from the jamming point. 
We will fix therefore the value of $\alpha$, and starting from the jamming point $\sigma_J$ we will increase the pressure entering the jammed phase.
We will consider only the case in which $\alpha>2$ so that the initial configuration is at a non-convex jamming transition point.
Therefore, compressing the system we expect, based on the phase diagram of Fig.\ref{phase_diagram}, that the system undergoes a transition form a glassy phase to the convex phase where the landscape reduces to a unique minimum.

\begin{figure}
  \centering
  \includegraphics[scale=0.7]{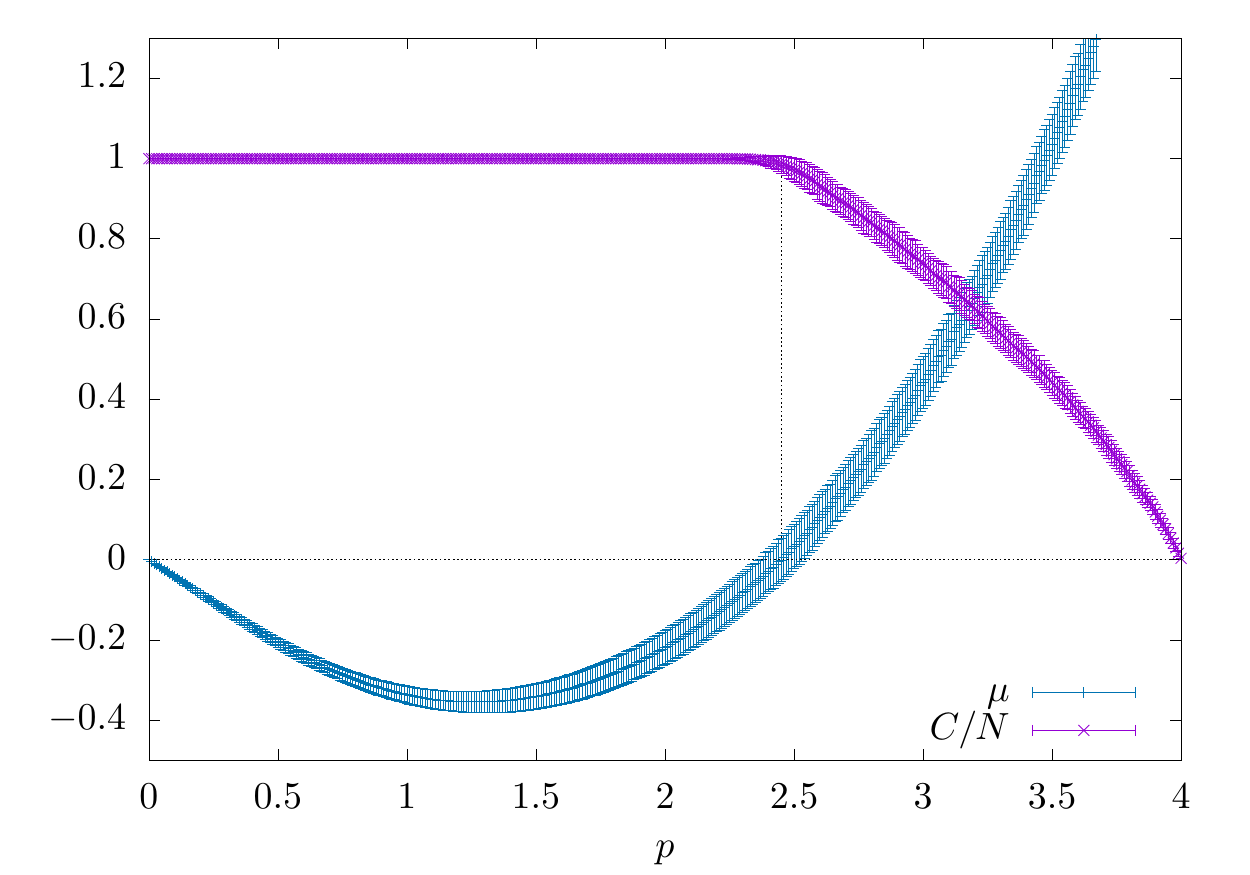}
  \includegraphics[scale=0.7]{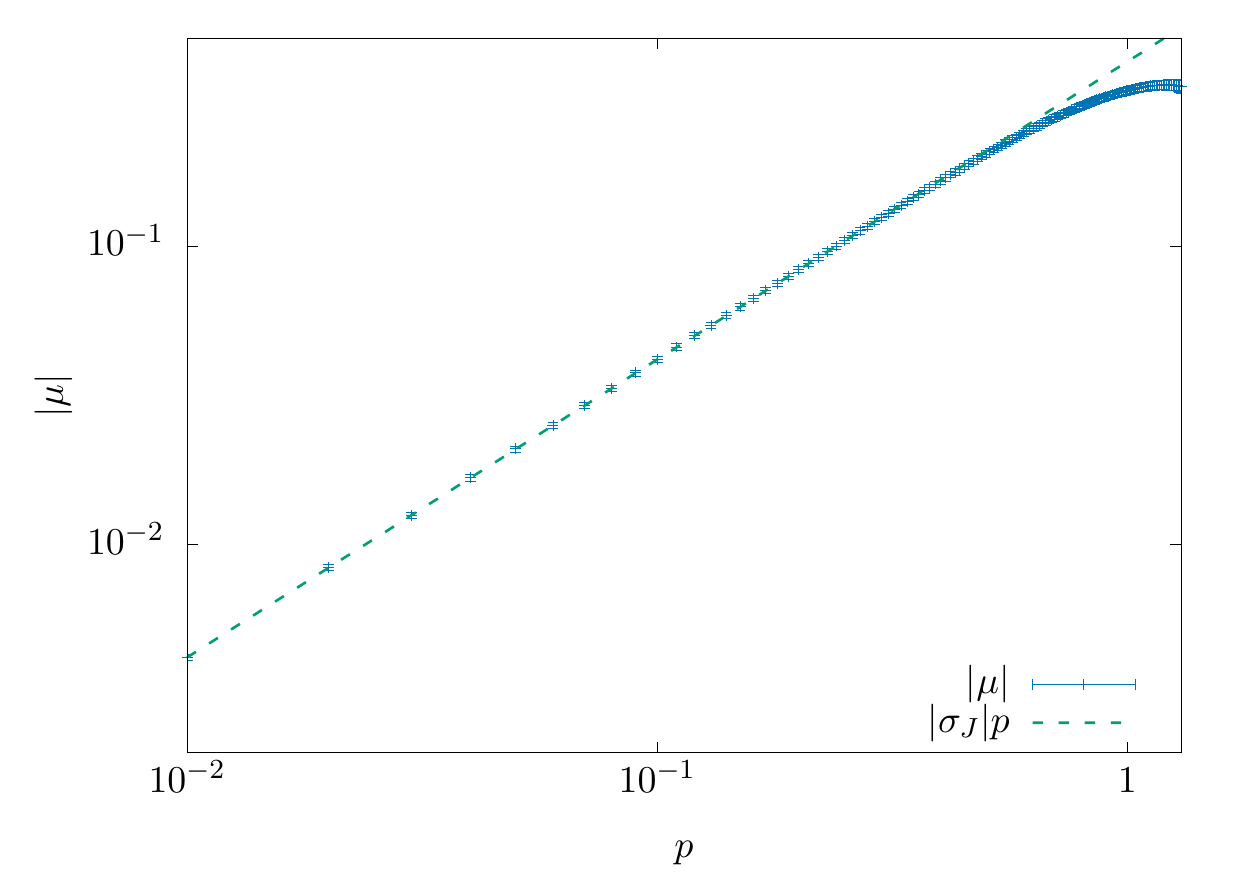}
  \includegraphics[scale=0.7]{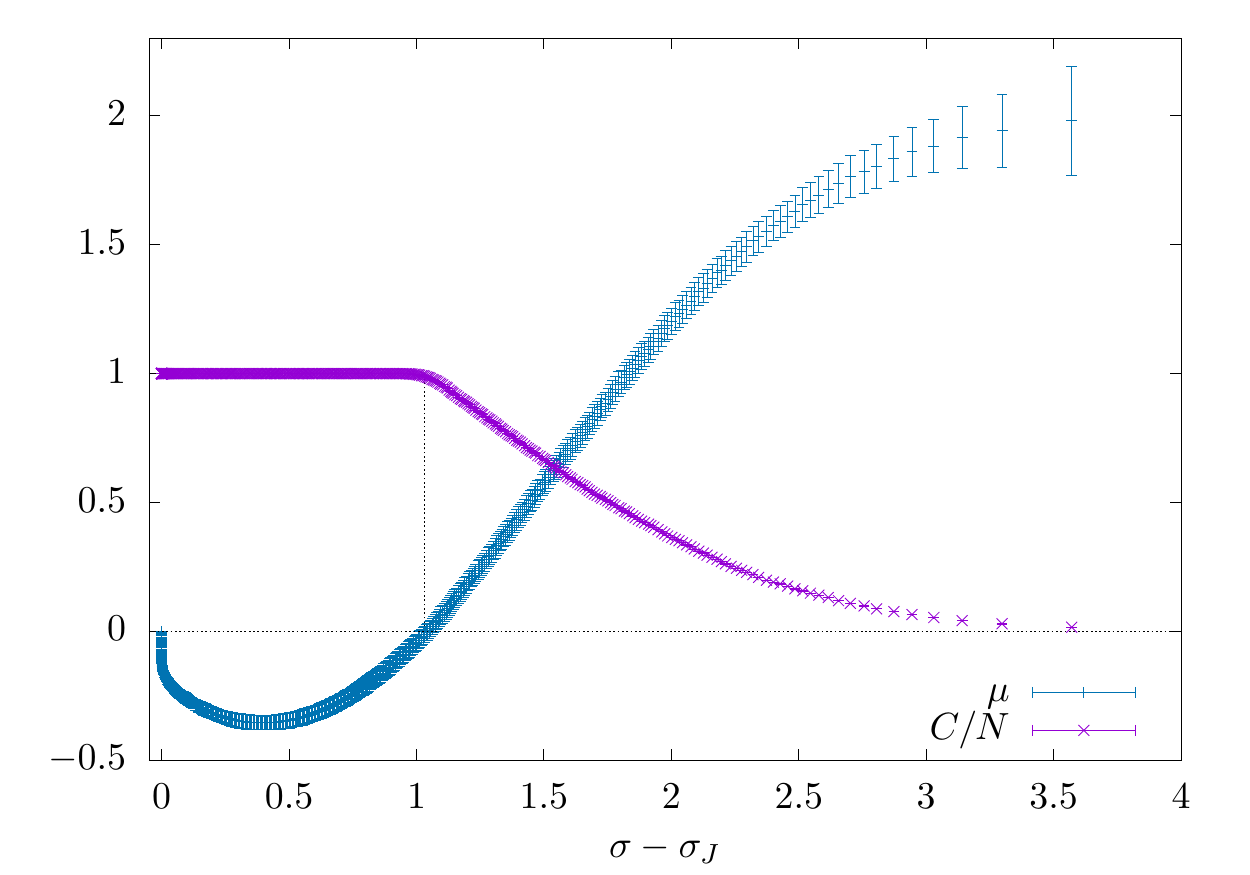}
  \includegraphics[scale=0.7]{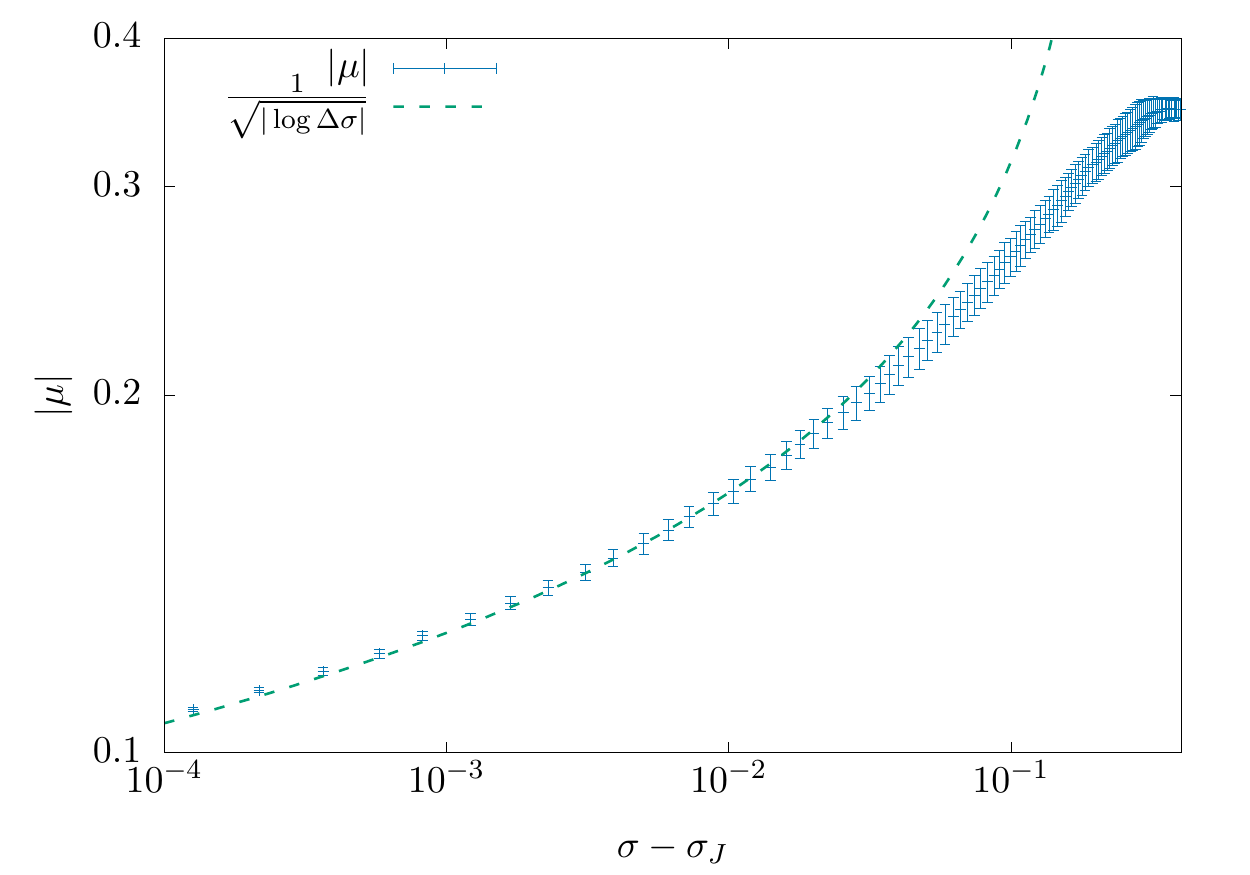}
  \caption{\emph{Left panel}: The evolution of the the number of contacts normalized by $N$ and the Lagrange multiplier $\lm$ as a function of $p$ (top plot) and $\sigma-\sigma_J$ (bottom plot) at $\alpha=4$ and $N=256$. The dotted lines correspond to the topology trivialization transition point (\emph{a.k.a.} the RSB transition) where the landscape changes from being glassy to being convex. \emph{Left panel}: Behavior of the absolute value of the Lagrange multiplier $\lm$ near the unjamming transition with respect to $p$ (top plot) and $\sigma-\sigma_J$ (bottom plot). We observe  a linear dependence on $\mu$ in $p$, $\mu\simeq \sigma_J p$, and a logarithmic dependence in $\sigma-\sigma_J$, see discussion of section \ref{sec:unJ}. }
  \label{mu_C}
  \end{figure}

In Fig.\ref{mu_C} we plot the evolution of both the number of contacts as well as the value of the Lagrange multiplier $\lm$ that we use to enforce the spherical constraint. 
We plot the corresponding behaviors both as a function of the pressure and as a function of the distance from the algorithmic jamming point $\sigma_J$.
It is clear from the figure that there are two regimes. 
For $p<p^*$ one has that the isostaticity index defined as $c=C/N$ is strictly equal to one. In this regime the system surfs on isostatic minima. Correspondingly the Lagrange multiplier $\lm$ is negative. This can be understood as follows. The cost function in Eq.~\eqref{lagra} is effectively linear in the degrees of freedom apart from the term proportional to $\lm$. Therefore the convex or non-convex nature of the problem is self-generated 
and mirrors in the sign of the Lagrange multiplier $\lm$.
The region where $\lm<0$ corresponds to the glassy phase where the optimization problem is non-convex while $\mu>0$ corresponds to the convex phase where the landscape is made of a unique attractive minimum. Therefore the replica symmetry breaking transition point 
at which $\mu$ changes sign is the point of a topology trivialization: it separates a region where the landscape is very rough and the dynamics surfs on marginally stable states from a region where the landscape is convex.

\subsection{Statistics of avalanches in the non-convex UNSAT phase}\label{sec_avalanches}
\begin{figure}
\centering
\includegraphics[width=0.6\columnwidth]{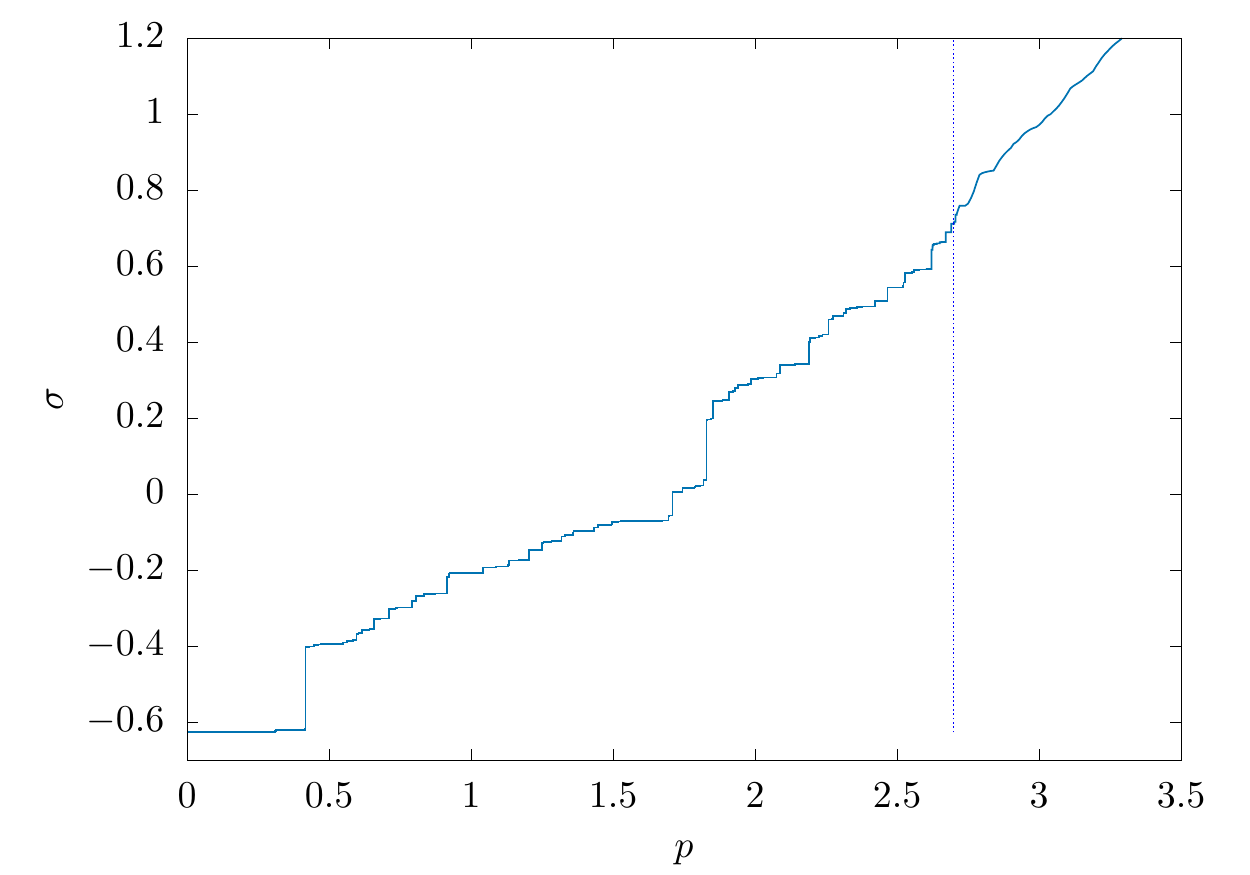}
\caption{The Devil staircase 
of the values of $\sigma$ as a function of $p$ for a single sample at $\alpha=4$ for $N=32$, starting from jamming. The vertical dashed line signals the point where $\mu$ changes signs and the landscape of the model becomes convex. Correspondingly, coming from jamming, one looses the jerky staircase profile that characterizes the non-convex phase and signals crackling noise.}
\label{staircase}
\end{figure}

In this section we consider the statistics of jumps when we follow the evolution of a local minimum upon increasing the pressure. 
Indeed, as we have already seen, as soon as the pressure increases
the system undergoes a series of avalanches, that are directly induced by the fact that some of the contact forces
may exit the support $(0,1)$ leading to a rearrangement of the contact network, with consequent jumps in $\sigma$ and energy
see Fig.\ref{staircase}.
In order to quantitatively describe the statistics of avalanches, 
we need to establish both the typical finite size amplitude of the variation of the pressure that
lead to a rearrangement, $\delta p$, as well as the 
one of the jumps of $\delta \sigma$ that take place when the rearrangements happen.

\subsubsection{The jumps of the pressure}

As we have seen, minima have a small range of stability when the pressure is changed. We would like to estimate here the order of magnitude of pressure changes $\delta p$ needed to destabilize a minimum. This is different at jamming and in the jammed phase. We have seen that at jamming, supposing
a Gaussian tail of the force distribution, $\delta p_J\sim 1/\sqrt{\log N}$. In the jammed phase, where we concentrate here, the situation is different. The force distribution has a pseudo-gap rather than an exponential tail, namely $\rho_f(f)\sim(1-f)^\theta$, and one can expect that typically $\delta p$ 
to behave as an inverse power of $N$, $\delta p\sim N^{-\beta}$.

In order to estimate the exponent $\beta$ we start from Eq.~\eqref{pe}. 
Since in the interval $[p,p+\delta p]$ the system does not move,
we have that
\begin{equation}
\delta \mu = \sigma\delta p 
\end{equation}
(remember that $\sigma$ does not change either before an avalanche).
Therefore the force balance condition, upon increasing pressure by $\delta p$ reads
\begin{eqnarray}
&&-v_i+ 
    \sum_{c} (f_{c}+\delta f_c) \frac{-\xi_{c,i}}{\sqrt{N}}+
    (\lm+\sigma\delta p )X_i = 0. \\
&&v_i=\sum_{o} \frac{\xi_{o,i}}{\sqrt{N}} \nonumber
\end{eqnarray}
or, defining $\tilde{\xi}_{c,i}=\frac{\xi_{c,i}}{\sqrt{N}}$
\begin{eqnarray}
  \label{eq:9}
\delta f_c=\delta p\sigma \sum_i (\tilde{\xi}^{-1})_{c,i}X_i.
\end{eqnarray}
We would like to argue that, while $\frac 1 N\sum_c \delta f_c = \delta p$, 
the typical values of $f_c$ are of order $O(\delta p\sqrt{N})$. 
 An increase of pressure results in a additional force on each variable $i$ proportional to $X_i$. These forces need to be compensated by variations in the contact forces $f_c$. These forces are correlated with the $\xi_c$ by the contact conditions.  In order to proceed, 
we would like to argue that the effect of a pressure change is statistically similar to the application of random forces
on the variables \footnote{In infinite dimension, it is clear that any direction is effectively a random direction from the perspective of the system. While it can be shown analytically for thermodynamic observables \cite{Ago20}, it is much less obvious for mesoscopic quantities, meaning quantities that scale in a nontrivial way with the system size as we are showing.}. Let us then slightly modify the problem, imagining to perturb the equilibrium equations by a random term $\epsilon Y_i$ with ${\bf Y}$ a vector of random variables independent of the patterns with $\langle  Y_i^2\rangle$=1. Since the resulting $\delta f_c$
will have both signs, we can estimate the order of magnitude of each term, by studying 
\begin{eqnarray}
  \label{eq:4}
  \frac 1 N \sum_c (\delta f_c)^2=\epsilon \frac 1 N \sum_{i,j} Y_i (M^{-1})_{ij} Y_j 
\end{eqnarray}
where the matrix $M_{ij}$ is 
\begin{eqnarray}
  \label{eq:10}
  M_{ij}=\frac 1 N \sum_c \xi_{c,i}\xi_{c,j}. 
\end{eqnarray}
This matrix is known to be 
 close to a 
Wishart matrix with quality factor $1$ \cite{FPUZ15}. In the thermodynamic limit its spectral density behaves as $\rho(\lambda)
\sim \lambda^{-1/2}$ for small $\lambda$, and, for finite $N$, the minimum eigenvalue is of order $\lambda_{min}\sim N^{-2}$. 
Moreover, the eigenvectors $|n\rangle$ of $M$ are just random points on the sphere of radius 1 independent from the eigenvalues,
orthogonal to each other and weakly correlated with $X$.
Let us use the spectral representation of $M$ and write:
\begin{eqnarray}
  \label{eq:11}
   \frac 1 N \sum_{i,j} Y_i (M^{-1})_{ij} Y_j =\frac 1 N \sum_n \frac {1}{\lambda_n} \langle Y|n\rangle^2.
\end{eqnarray}
The random factors $\langle Y|n\rangle$ are Gaussian variables with unit variance, so that, the expected value over $Y$ of this quantity is
\begin{eqnarray}
  \label{eq:13}
  \frac 1 N \;{\rm Tr}\; M^{-1}=\int_{\lambda_{min}}^4 \frac{\rho(\lambda)}{\lambda}\sim\lambda_{min}^{-1/2}\sim N,
\end{eqnarray}
leading to $\delta f_c=O( \epsilon N^{1/2})$. We verified that the calculation of higher moments confirms this scaling. 
We can go now back to the original perturbation proportional to $X$. 
Here the situation is more subtle as we have that 
\begin{equation}
  \label{eq:7}
  f_c=\sum_i  (\tilde{\xi}^{-1})_{c,i} v_i +\mu \sum_i  (\tilde{\xi}^{-1})_{c,i}  X_i \in (0,1). 
\end{equation}
We notice now that the contribution from the overlaps $v_i=\sum_o \tilde{\xi}_{o,i}$  is a random term essentially independent of the choice of the 
contacts. If $O\sim N$, $v_i$ is of order $O(1)$ and the result of  (\ref{eq:7}) must come from the cancellation of terms both of $O(\sqrt{N})$. 

We conclude therefore that, $\delta f_c$, which is proportional to the second term in the r.h.s. of (\ref{eq:7}),  is
 $\delta f_c=O(\delta p N^{1/2})$. This scaling does not apply close to jamming,  there $|v_i|\ll |X_i|$ and there is no cancellation,  
the second contribution in (\ref{eq:7}) dominates the sum and $\delta f_c=\delta p f_c$.  
We notice that this argument has an interesting byproduct: it shows that in the bulk of the jammed phase, the effect of a 
small compression is statistically equivalent to a random perturbation. Some implications of this observation will be discussed in the conclusions. 

We can now estimate the order of magnitude of a destabilizing pressure variation $\delta p$. 
Following a compression, the first force that exits the stability support $(0,1)$ is one close to the edges in the unperturbed configuration; standard extreme statistics tells us that the corresponding $\delta f$ is of the order $\delta f_c\sim N^{-1/(1+\theta)}$. We therefore obtain the scaling
\begin{eqnarray}
  \label{eq:12}
   N^{-1/(1+\theta)}\sim N^{1/2} \delta p
\end{eqnarray}
or, 
\begin{equation}
\delta p\sim N^{-\beta} \ \ \ \ \ \ \ \ \ \ \ \ \beta=\frac 12 +\frac {1}{1+\theta}
\label{scal_dp}
\end{equation}

\begin{figure}
	\centering
	\includegraphics[width=0.47\columnwidth]{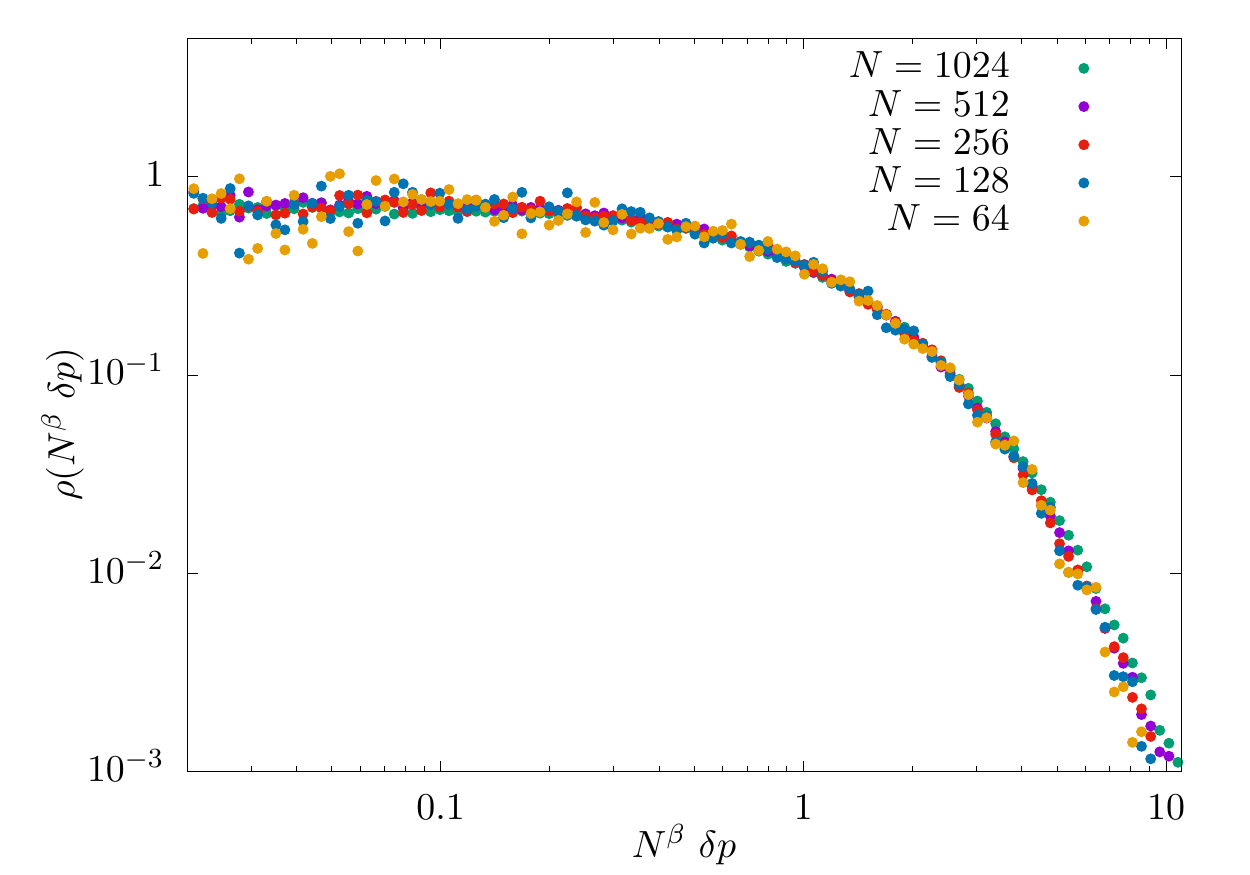}
	\includegraphics[width=0.47\columnwidth]{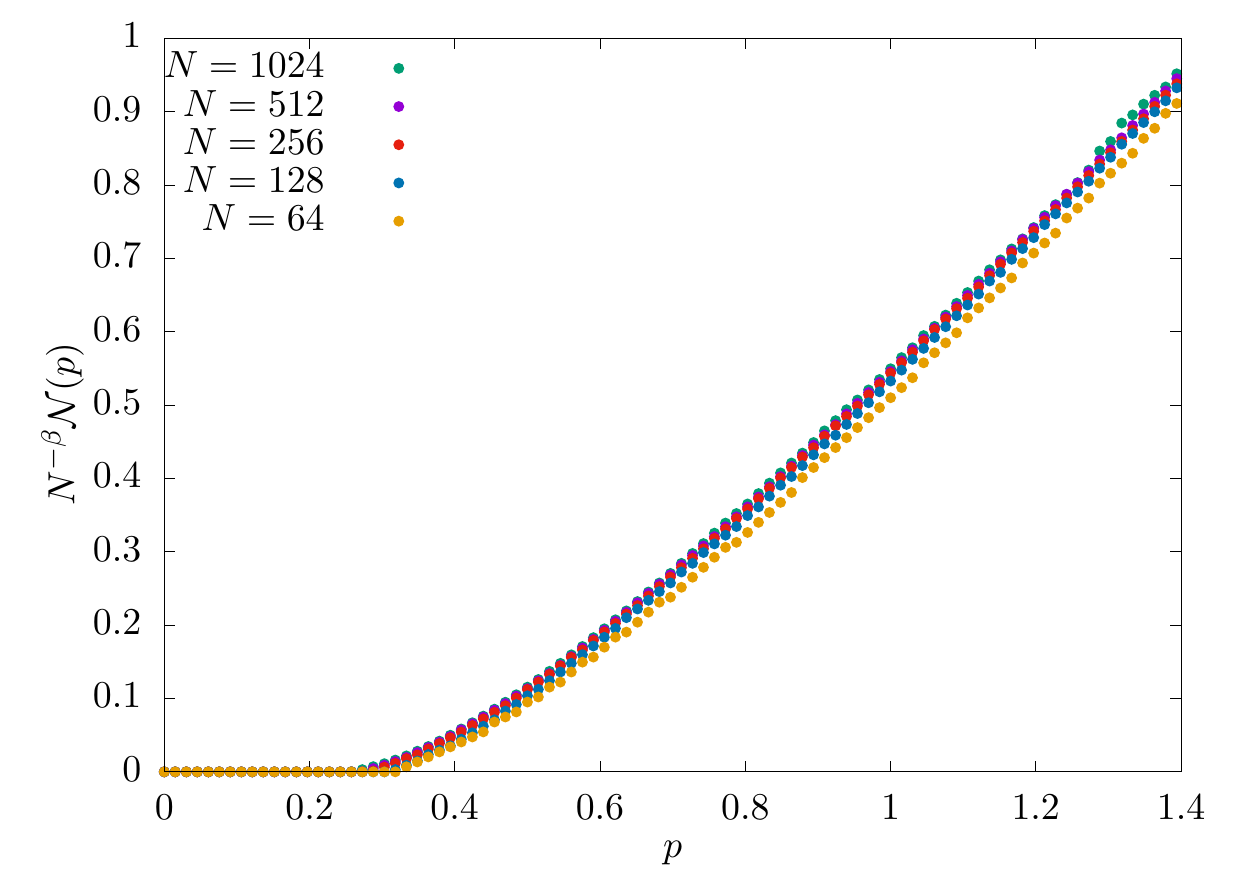}
\caption{\emph{Left panel:} Statistics of the scaled jumps of the pressure $\hat{\delta p} = N^{\beta}\delta p$. \emph{Right panel:} Average cumulative number of plastic events up to pressure $p$, rescaled by $N^{-\beta}$,  as a function of $p$. It shows that the number of plastic events scales as $N^\beta$ for finite pressure variation $\Delta p$.}
\label{PRESSIONE}
\end{figure}

In Fig.\ref{PRESSIONE}, we show the histogram of the rescaled pressure jumps $\delta \hat{p}= N^\beta \delta p$ between plastic events, 
collected from all the jumps taking place when the pressure lies in the interval $p\in [ 0.2, 1.2]$\footnote{We verified that in this interval 
of pressure the statistics of jumps is reasonably stationary.} for different system sizes and we observe an excellent data collapse.
Notice that the tail of the distribution at large argument seem to converge to zero exponentially and in any case much faster than $(\delta\hat{p})^{-2}$.
The first moment of $\delta \hat{p}$,  ${\langle \delta \hat{p}\rangle}$ remains finite for $N\to\infty$, and that the number ${\cal N}(\Delta p)$
of plastic event that occur when pressure is increased by a small but finite amount $\Delta p$ scales as,
\begin{eqnarray}
\label{eq:17}
{\cal N}(\Delta p)=\frac{\Delta p}{\langle \delta p\rangle}\sim \Delta p N^\beta
\end{eqnarray}

\begin{figure}[h]
\centering
\includegraphics[width=0.47\columnwidth]{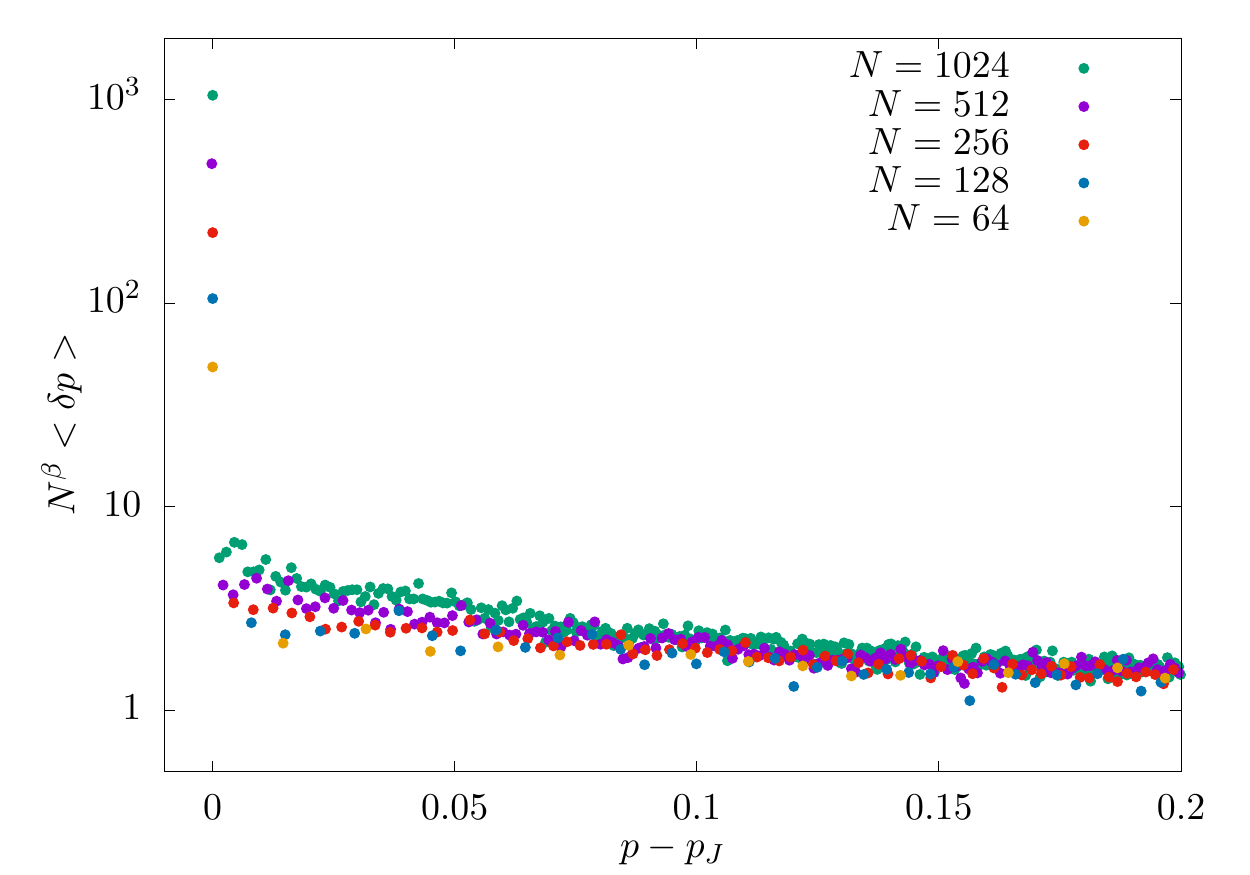}
\includegraphics[width=0.47\columnwidth]{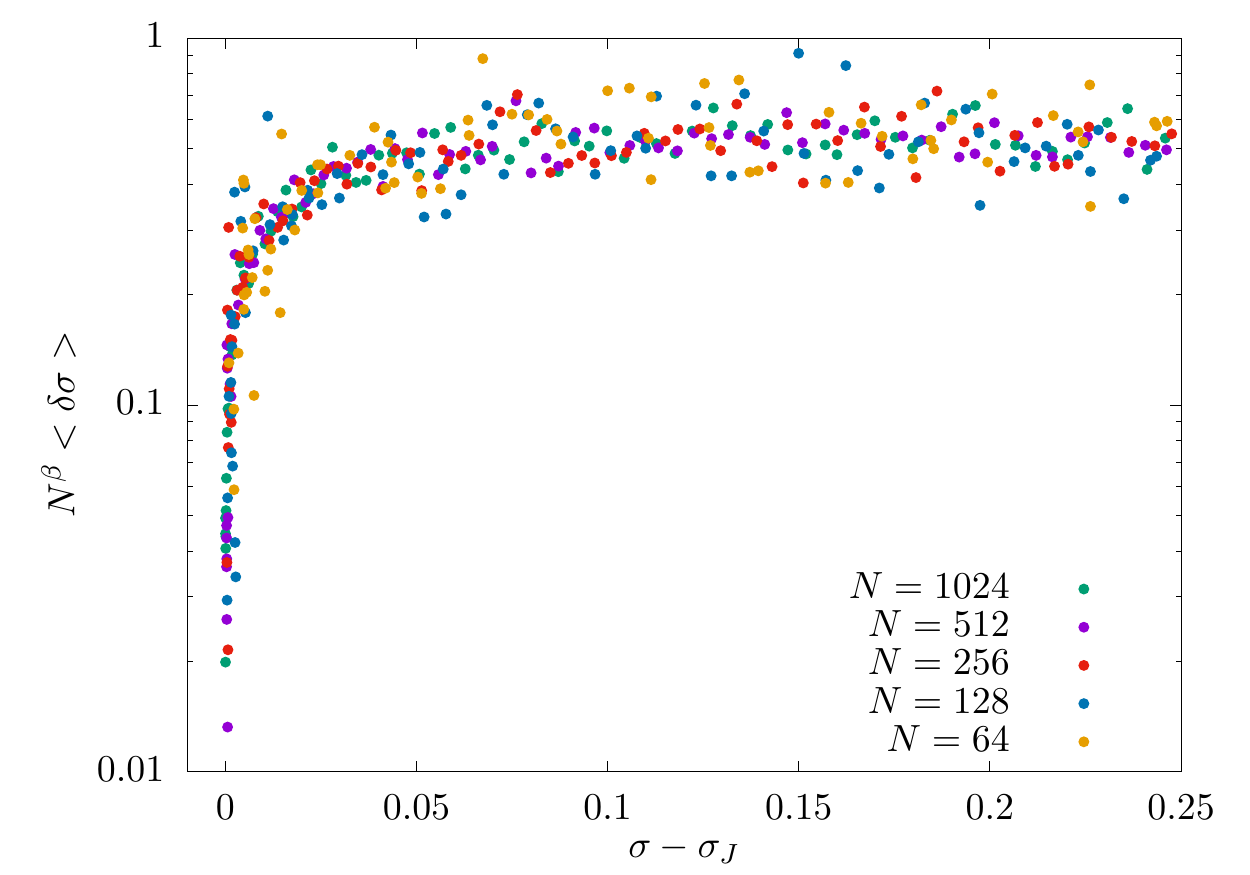}
\caption{{\it Left panel:} Average jump $\langle\delta p\rangle$, rescaled by $N^\beta$ as a function of $p-p_J(N)$ for $N=64,128,256,512,1024$. 
The first point is the pressure jump at jamming $p_J\sim \frac{1}{\sqrt{\log N}}$. All the subsequent jumps are much smaller. Notice that the scaling 
$\langle\delta p\rangle\sim N^{-\beta}$ is very well verified far away from jamming. 
{Close to jamming there are small deviations to this behavior. It is not clear to us if these are due to next to leading 
finite-size corrections or to genuine changes in the leading behavior. }
{\it Right Panel:} Average rescaled avalanche size $\langle \delta \sigma \rangle N^\beta$ as a function of $\Delta \sigma=\sigma-\sigma_J$ for the same values of $N$. The scaling is well respected till the jamming point.}
 \label{fig:deltap}
\end{figure}

The left panel of Fig. \ref{fig:deltap} shows the average $\langle \delta p\rangle$ as a function of pressure. As expected the scaling $\langle \delta p\rangle\sim N^{-\beta}$ is well respected except for the vicinity of jamming. 

\subsubsection{Statistics of the jumps of $\sigma$}
In a similar way we can relate the order of magnitude of the jumps in $\sigma$ after a plastic event to the exponent $\gamma$ of the gap distribution.  
For $N\to \infty$, the statistics of the small jumps of $\sigma$ shrinks to zero as $\delta \sigma \sim N^{-\omega}$.
To determine $\omega$, let us consider the variation of $\sigma$ that ensue a process where a weak single contact, say contact $c$, 
becomes a small gap or overlap. 
Therefore we have that for the special contact $c$ that has been opened
\begin{equation}
\sum_i {\tilde{\xi}_{c',i} \delta X_i} - \delta \sigma=
  \delta_{c,c'} \delta h_c
\end{equation}
where using again extreme statistics we can estimate the magnitude of the smallest gaps $\delta h_c\sim N^{-1/(1-\gamma)}$. 
Inverting such relation we get
\begin{equation}
\delta X_i =  \sum_{c'} {\tilde\xi}^{-1}_{c',i} \left(  \delta_{c,c'} \delta h_c+\delta \sigma\right)
\end{equation}
Multiplying both sides by $X_i$ 
\begin{eqnarray}
  \label{eq:14}
  X\cdot\delta X=-\frac 1 2 \delta  X\cdot\delta X=\frac{1}{\sigma \delta p}
\left(
\delta f_c\delta h_c+\delta\sigma \sum_{c'}\delta f_{c'}.  
\right)
\end{eqnarray}
If we impose that the two terms in the r.h.s. of (\ref{eq:14}) are of the same order of magnitude, much larger than the one of l.h.s, the we get
\begin{equation}
\label{scale_ds_1}
\delta \sigma \sim  N^{-\frac 1 2 -\frac{1}{1-\gamma}} \ \ \ \ \ \ \ \ \ \ \omega=\frac 12 +\frac{1}{1-\gamma}
\end{equation}
Using the scaling relation $\gamma=(2+\theta)^{-1}$ \cite{Wy12} we get
\begin{equation}
\omega= \frac 32 +\frac{1}{1+\theta}=1+\beta\:.
\label{scale_ds}
\end{equation}
% In a similar way, we can estimate the typical steps of the square displacement $(\delta X)^2$ to scale as 
% \begin{eqnarray}
%   \label{eq:15}
%   \frac{1}{N} (\delta X)^2\sim N^{-2/(1-\gamma)}=N^{-(1+2\beta)}. 
% \end{eqnarray}
Our argument could be extended to compression of jammed configurations of soft linear spheres giving rise to identical exponents. In fact, we 
notice that both Eq.~\eqref{scal_dp} and Eq.~\eqref{scale_ds} coincide with the ones obtained for hard spheres at jamming under shear strain \cite{CR02,DDLW15}. This indicates that any destabilizing perturbation that leads isostatic states to new isostatic states with self-similar distributions of forces and gaps close to the edges gives rise to the same kind of avalanche statistics. 
\begin{figure}[h]
\centering
\includegraphics[width=0.47\columnwidth]{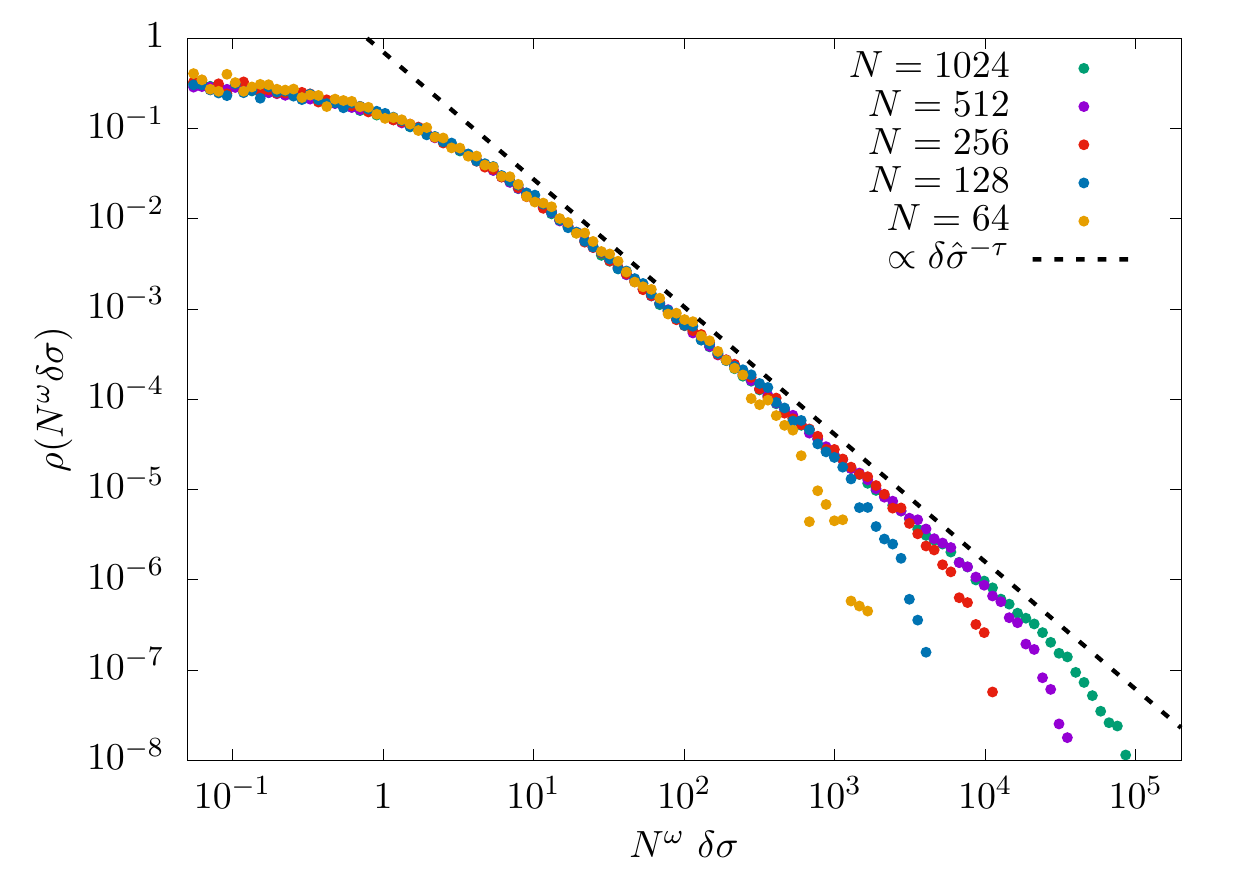}
\includegraphics[width=0.47\columnwidth]{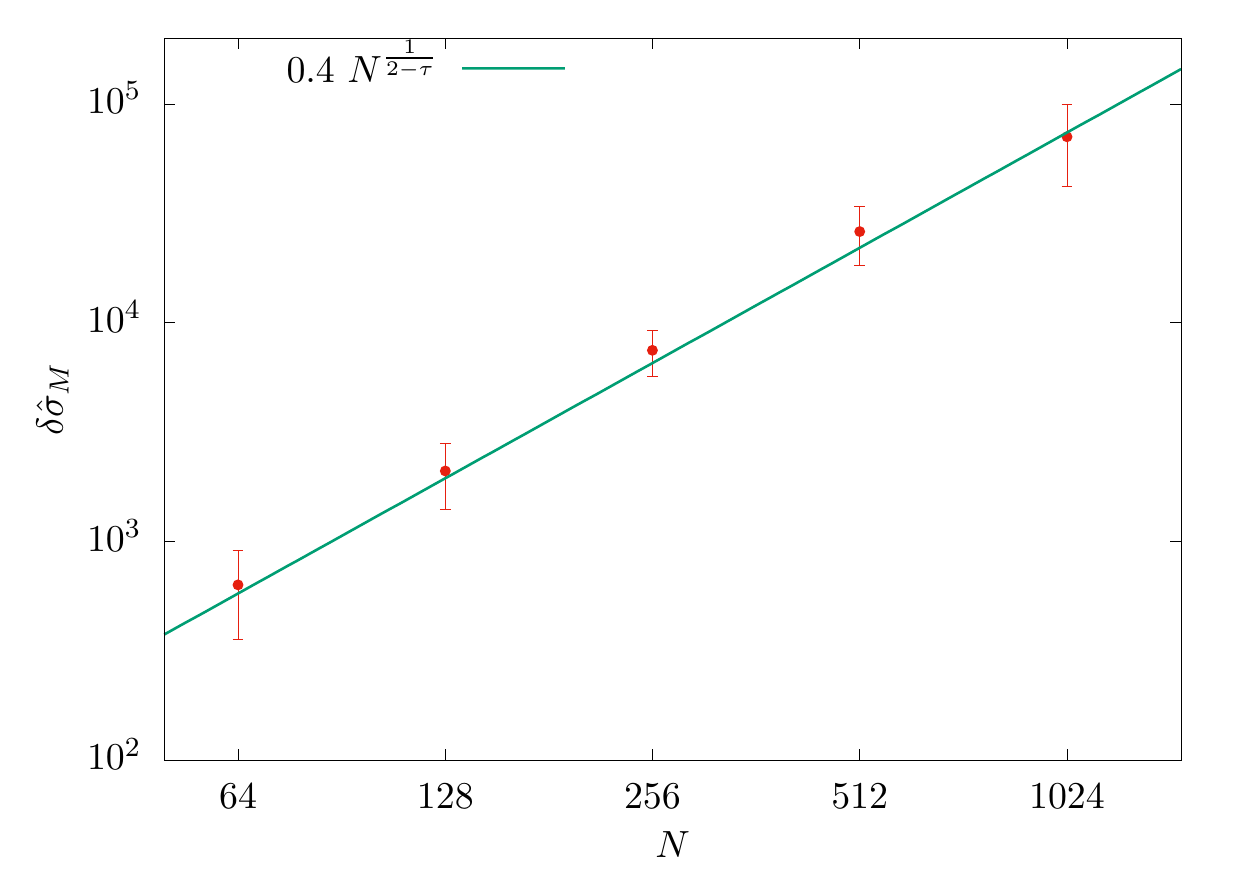}
\caption{\emph{Left Panel}. Main plot: statistics of the scaled avalanche size $\hat{\delta \sigma} = N^{\omega}\delta \sigma$. The dotted line the power law $\rho(\hat{\delta \sigma})\sim \hat{\delta \sigma}^{-\tau}$. \emph{Right Panel}. The scaling of $\hat{\delta\sigma}_M$ as a function of the system size. The green dashed line is a fit to $\hat{\delta \sigma}_M\sim N^{\frac{1}{2-\tau}}$.} \label{Ava-fig}
\end{figure}
Notice that the scaling of (\ref{scale_ds_1}) is incompatible with a distribution of the rescaled jumps $\delta\hat{\sigma}$ which admits a finite first moment in the thermodynamic limit. In fact, a finite pressure increase $\Delta p={\cal N}\langle \delta p\rangle$, should correspond to a finite jump $\Delta \sigma = {\cal N}\langle \delta \sigma\rangle$. This tells that  $\langle \delta \sigma\rangle=N^{-\omega} 
\langle \delta \hat{\sigma}\rangle \sim N^{-\beta}$ and that the finite $N$ average of 
$\delta \hat{\sigma}$ should diverge as $\langle \delta \hat{\sigma}\rangle \sim N$ for large $N$.  
{As it can be seen in the right panel of Fig. 
\ref{fig:deltap} the scaling $\langle \delta\sigma\rangle\sim N^{-\beta}$ is observed already right after jamming.} 
The divergence of the first moment indicates that the distribution of avalanches 
$\delta \hat{\sigma}$ should exhibit in the thermodynamic limit a power law at large argument 
\begin{eqnarray}
  \label{eq:18}
  \rho(\delta\hat{\sigma})\sim \delta\hat{\sigma}^{-\tau} \ \ \ \ \ \delta\hat{\sigma} \gg 1. 
\end{eqnarray}
The exponent $\tau$ should be in the interval $1<\tau \le 2$, such that the
the distribution has a divergent first moment. For finite $N$ however, the distribution should be cut-off around a value $\delta \hat{\sigma}_{M}$ so that 
\begin{eqnarray}
  \label{eq:20}
 \langle \delta \hat{\sigma}\rangle\sim  
\int^{\hat{\delta\sigma}_M}_0 d\hat{\delta\sigma} (\hat{\delta\sigma})^{1-\tau} \sim  (\hat{\delta\sigma}_M)^{2-\tau}
\sim N.
\end{eqnarray}
It follows that
\begin{equation}
\hat{\delta\sigma}_M\sim N^{\frac{1}{2-\tau}}
\end{equation}
The statistics of avalanches in mean-field disordered systems has been fully characterized using equilibrium techniques \cite{LMW10,FS17}, where instead of studying the change of local minima following a destabilizing perturbation, one studies the discontinuities in the evolution of the actual ground state of the system when this is in a fullRSB region.  In this case, the exponent $\tau$ can be related to the force pseudogap exponent by the relation
\begin{eqnarray}
  \label{eq:19}
\tau=\frac{3+\theta}{2+\theta}\simeq 1.41. 
\end{eqnarray}
Differently from the scaling with $N$, that we have obtained by purely local considerations,  this form for the 
avalanche distribution with the specific value of $\tau$ depends on the statistical properties of the neighborhood of the ground state, something that is captured by the replica solution. Remarkably we find that within numerical precision, the value (\ref{eq:19}) coincides with the one found for avalanches of sheared hard spheres at jamming \cite{CR00}
and soft spheres close to jamming \cite{FS17}.  Our simulations indicate that the remarkable coincidence between static and dynamic avalanche statistics also holds in this case.  In Fig.\ref{Ava-fig}-Left Panel we show the statistics of the rescaled jumps of $\hat{\delta \sigma} = N^{\omega}\delta \sigma$ collected from all the plastic events taking place when the pressure lies in the interval $ [ 0.2, 1.2]$ when a minimum at jamming is followed upon compression. 
For small enough size the numerical results collapse onto each other and approach the expected power law distribution (\ref{eq:18}). In the right panel 
we show that the maximum $\delta\hat{\sigma}_M$ around which the power law is cut-off respects the expected scaling with $N$.  
\begin{figure}[h]
	\centering
	\includegraphics[width=0.47\columnwidth]{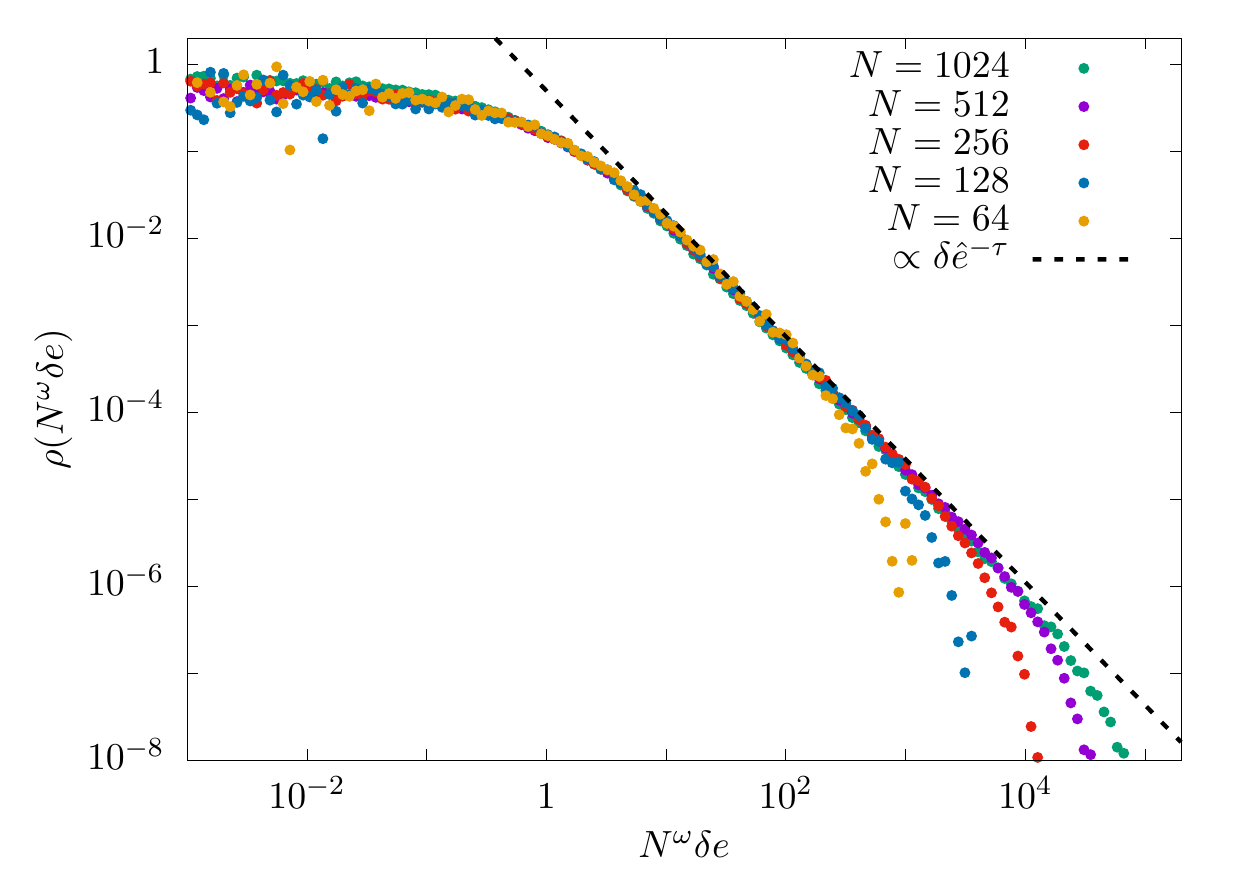}
	\includegraphics[width=0.47\columnwidth]{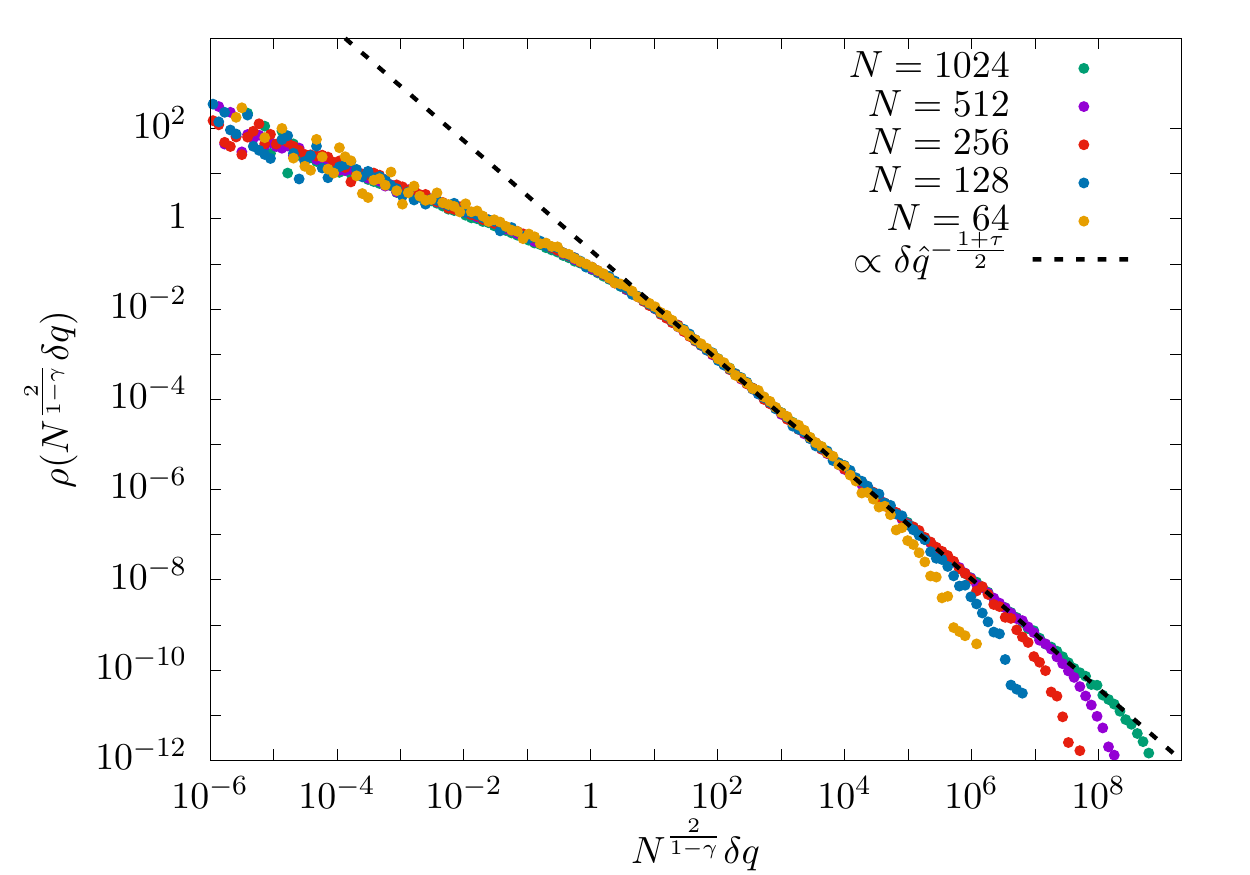}
	\caption{\emph{Left Panel}. Statistics of the scaled energy variations $\hat{\delta e} = N^{3/2+1/(1+\theta)}\delta e$. The dotted line is a  fit to the form $\rho(\hat{\delta e})\sim \hat{\delta e}^{-\tau}$.
\emph{Right Panel}. Statistics of the scaled squared displacements $\hat{\delta q} = N^{2/(1-\gamma)}\delta q$. The dotted line is a fit to  $\rho(\hat{\delta q})\sim \hat{\delta q}^{-\frac{\tau+1}{2}}$. } 
\label {fig:eq}
\end{figure}

All in all the results of Fig.\ref{Ava-fig} strengthen the observation that non-convex jammed phase of the linear perceptron is marginally stable, self-organized critical and belongs to the jamming universality class. 

\subsection{Energy and Overlap avalanches}

A similar analysis as in the previous section allows to analyze the jumps in energy, and 
in position. It is easy to see that the jumps in energy density follow exactly the same scaling as the ones in $\sigma$, namely $\delta e\sim N^{-\omega}$. As in the case of $\sigma$ a power law avalanche distribution with divergent first moment follows, and it is possible to see that  the static avalanche exponent coincides  with $\tau$ 

As far as the jumps in positions are concerned, we can estimate the typical steps of the square displacement $(\delta X)^2$ to scale as 
\begin{eqnarray}
  \label{eq:15}
\delta q=  \frac{1}{N} (\delta X)^2\sim N^{-2/(1-\gamma)}=N^{-(1+2\beta)}. 
\end{eqnarray}
Also in that case we have a power law distribution of jumps. The static avalanche exponent  \cite{LMW10,FS17}, is in this case $\tau'=\frac{\tau+1}{2}$. 
In Fig. \ref{fig:eq} we display the probability distributions of energy and displacement jumps, which confirm the above scaling, and shows that also in this case the dynamical 
avalanche exponents coincide with the static ones. 

\begin{figure}
	\centering
	\includegraphics[scale=0.7]{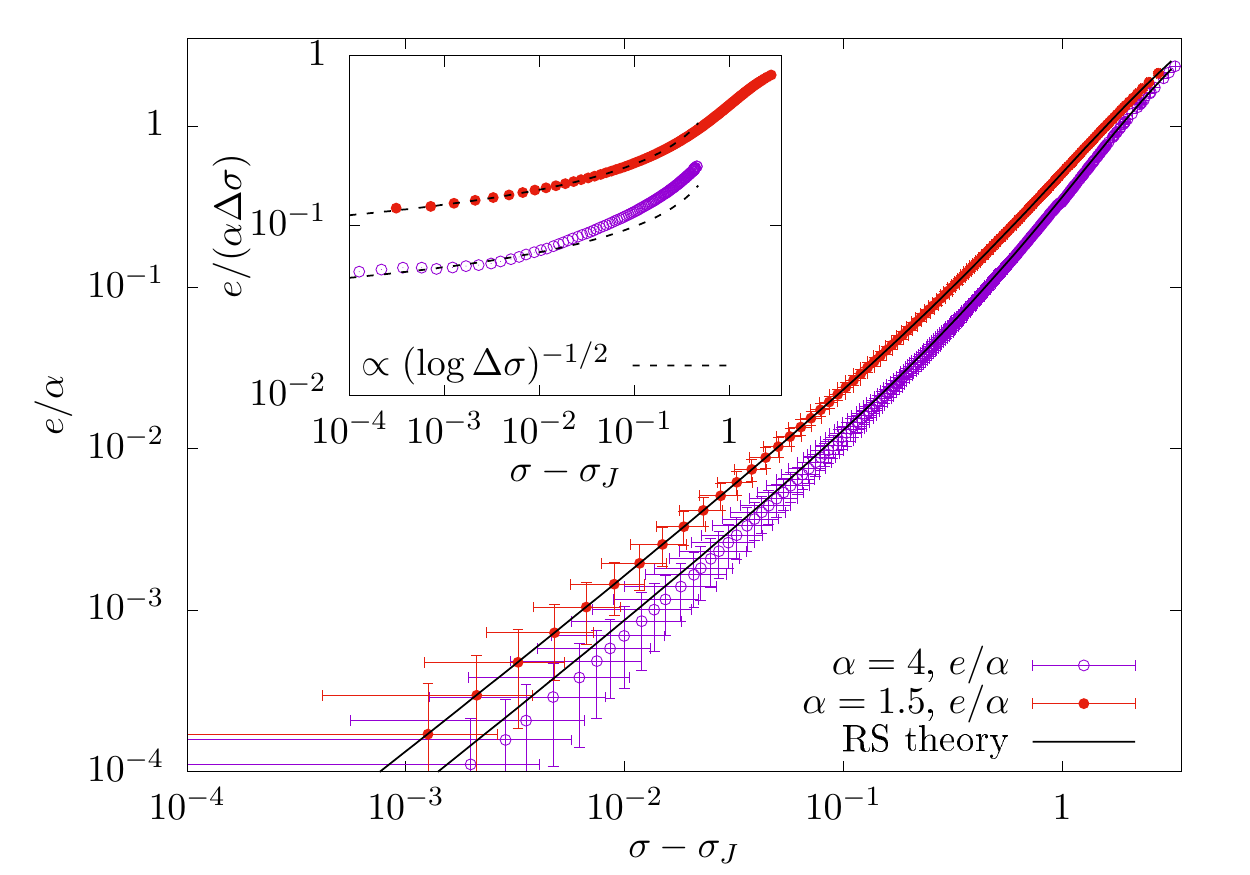}
	\includegraphics[scale=0.7]{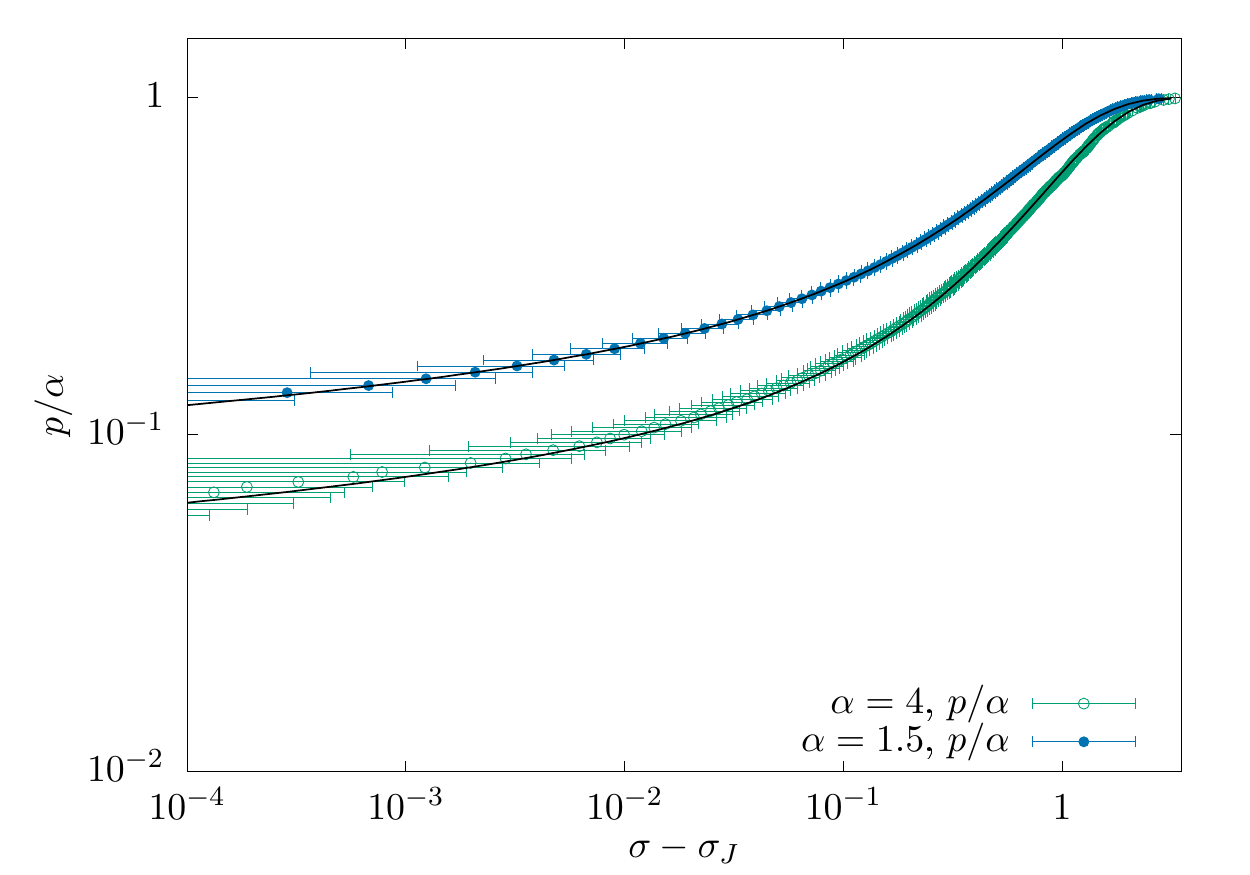}
	\caption{\emph{Left Panel}. The behavior of the scaled energy $e/\alpha$ as a function of the distance from jamming $\Delta \sigma=\sigma-\sigma_J$. 
		For $\alpha=1.5$, jamming is in the convex phase and we can use the replica symmetric theory to study the corresponding scaling behavior as in Eq.~\eqref{scalingRS}. For $\alpha=4$, jamming is in the non-convex region and in principle the replica symmetric theory is not valid anymore. Anyway, we fit the numerical curves with the replica symmetric approximation and we see a good agreement, signaling the fact that for the energy the scaling behavior as a function of the distance from jamming is preserved. In the inset we plot the same quantities divided by $\sigma-\sigma_J$. This way we reveal the presence of logarithmic corrections to the linear scaling of the energy with respect to $\sigma-\sigma_J$.  \emph{Right Panel}. The behavior of the scaled pressure $p/\alpha$ as a function of the distance from jamming $\Delta \sigma=\sigma-\sigma_J$. As for the energy, the replica symmetric theory gives a good prediction for the pressure also in the non-convex region. Data obtained with $N=2048$ for $\alpha=1.5$, $N=512$ ($N=1024$ in the inset) for $\alpha=4$. Error bars are sample to sample fluctuations.}  
	\label{scaling_J}
\end{figure}

\begin{figure}
	\centering
	\includegraphics[scale=1]{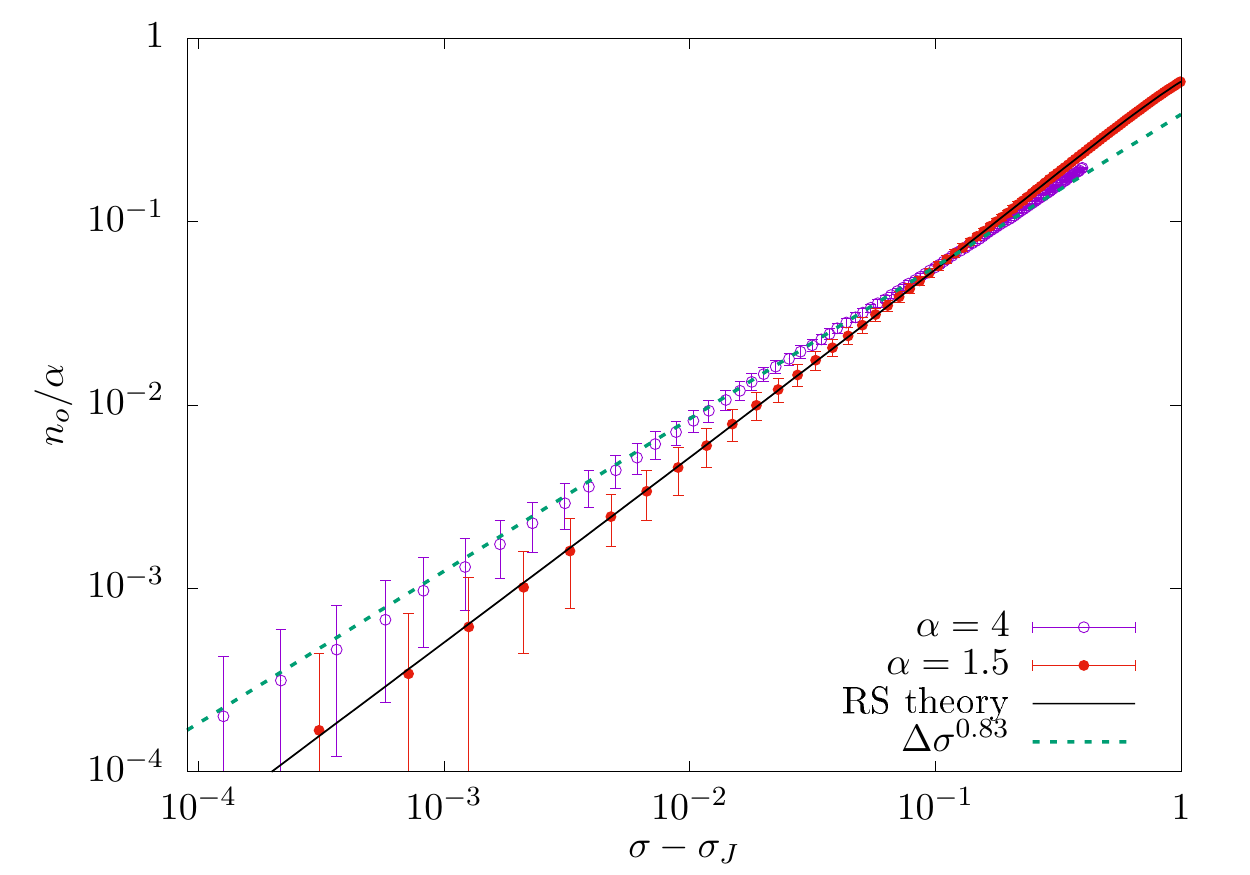}
	\caption{The scaled number of overlaps $n_o/\alpha$ as a function of the distance from jamming. In this case, the replica symmetric prediction clearly holds only in the convex region while the non-convex one shows the appearance of a non-trivial power law behavior. The dotted line represent the prediction form the scaling analysis. Data has been produced with $N=2048$ for $\alpha=1.5$ and $N=1024$ for $\alpha=4$. Error bars are sample to sample fluctuations.}
\label{fig:no}
\end{figure}

\section{The unjamming transition}
\label{sec:unJ}
In this section we study the unjamming transition, that occurs when pressure vanishes from positive values $p\to 0^+$. 
From the study of unjamming in soft spheres, the features of this transition when the exponent $a$ in the interaction potential is larger
than unity are  known {\cite{OSLN03, WSNW05,GLS16}.  The analysis can be extended to the perceptron along the line 
of  \cite{FPSUZ17} for the harmonic case. 
Using, as customary, the distance from the jamming point $\Delta \sigma=\sigma-\sigma_J$ as a control parameter
the pressure and the energy close to the transition behave as
 \begin{eqnarray}
  \label{eq:3}
&&  p\sim \Delta \sigma^{a-1}\\
&&  E\sim \Delta \sigma^a 
\end{eqnarray}
 while the variation in the density number of overlaps $\Delta z$ with respect to the isostatic value $z=1$, shows a square root singularity
 independently of $a$ 
 \begin{eqnarray}
   \label{eq:22}
   \Delta z\sim\sqrt{\Delta \sigma}. 
 \end{eqnarray}
The Lagrange multiplier $\mu$, according to eq. (\ref{pe}) is dominated by the pressure variation $\mu\approx \sigma_J\  p\sim \Delta \sigma^{a-1}$. 
The laws (\ref{eq:3}) are a consequence of the fact that for $a>1$,  an increase of the margin $\Delta \sigma$ causes  all the contacts to become overlaps 
with $|h_c|\sim \Delta\sigma$, 
independently of $a$, \cite{WSNW05,GLS16}.  Simple dimensional analysis gives then (\ref{eq:3}). In the compressed phase the leading excitations are linear. Relation (\ref{eq:22}) expresses a condition of stability for the linear modes \cite{WSNW05}. 

These relation should break down in the linear potential case $a=1$, as it manifest from the facts that (1) the pressure has to vanish for $\Delta\sigma\to 0$ and (2) all excitations are non-linear even away from jamming. 
The vanishing of the exponent relating $p$ to $\Delta \sigma$ suggests that a logarithmic behavior could appear
\begin{eqnarray}
  \label{eq:6}
&&  p\sim  1/\log(1/\Delta \sigma)^b \\
&&  E\sim \Delta \sigma/\log(1/\Delta \sigma)^b 
\end{eqnarray}
with $b$ a positive exponent. Again $\mu\approx \sigma_J p\sim  1/\log(1/\Delta \sigma)^b$.

The behavior of the different quantities close to jamming should be in principle accessible from the analysis of the exact mean-field equations of the replica method \cite{FSU19}.  This analysis is rather simple for $\alpha<2$, where replica symmetry holds. In that case, we get that close to jamming the pressure, energy, and density of overlaps behave as\footnote{more precisely, we can determine the leading behavior as: 
\begin{equation}
\begin{split}
e&\simeq \alpha \frac{1+\tau_J}{\sqrt{2\pi}} \frac{\Delta\sigma}{\sqrt{2|\ln((1+\tau_J)\Delta\sigma)|}}\ \ \ \ \ \ \ p\simeq\alpha \frac{2^{-\sigma_J^2/2}+\tau_J}{\sqrt{2\pi}}\frac 1{\sqrt{2|\ln((1+\tau_J)\Delta\sigma)|}}\\
n_o&=\frac ON \simeq \alpha \frac{1+\tau_J}{\sqrt{2\pi}}\Delta\sigma\ \ \ \ \ \ \tau_J=\sigma_J\sqrt{\frac\pi 2}  \left(1+\textrm{erf}\left(\frac{\sigma_J}{\sqrt 2}\right)\right).
\end{split}
\end{equation}
}

\begin{equation}
\begin{split}
e&\sim \frac{\Delta\sigma}{\sqrt{|\ln(\Delta\sigma)|}}\ \ \ \ \ p\sim\frac 1{\sqrt{|\ln((\Delta\sigma)|}}\ \ \ \ \ \ n_o=\frac ON \sim 
\Delta\sigma\\
\label{scalingRS}
\end{split}
\end{equation}

In Fig.\ref{scaling_J}-Left Panel we plot such scaling for $\alpha=1.5$ showing a good agreement. The predictions 
(\ref{scalingRS})
could be questioned in the non-convex case. Notice that in the case $a>1$, the replica symmetric analysis 
gives the correct scaling (\ref{eq:3}) of energy and pressure, while it predicts a linear behavior $\Delta z\sim \Delta\sigma$, rather 
than the square-root of the non-convex case.

In our case, if we look at the number of overlaps $n_0$ in numerical simulations, we find in the non-convex case we 
do not find either linear nor square root behavior.  In Fig.\ref{fig:no} we plot $n_o$ \emph{vs} $\Delta\sigma$ close to jamming for $\alpha=4$ in double-log scale. We observe power law behavior $n_o\sim \Delta\sigma^\nu$, with an an exponent $\nu$ smaller than one but larger than $1/2$, 
that we estimate $\nu \simeq .83$, compatible with the value $\nu =1/\beta$. The origin of $1/\beta$ can be traced to the behavior with $N$, of $\delta n_0$ and $\delta\sigma$ in avalanches close to jamming. Typical avalanches produce there a small number of contacts and $\delta n_o\sim 1/N$.  The statistics of jumps in $\sigma$ 
on the other hand is likely to give $\langle \delta \sigma\rangle N^{-\beta}$ till jamming. If we suppose that the scaling behavior for $\delta n_o\sim 1/N$ remains valid for small but finite $\delta n_o$ we find 
 \begin{eqnarray}
   \label{eq:23}
   \Delta n_o\sim \Delta \sigma^{1/\beta}. 
 \end{eqnarray}
The appearance of the logarithms in the behavior of pressure and energy can also be rationalized by qualitative scaling. The destabilizing jumps in pressure depend on the 
tail of the distribution of forces, which for the perceptron close to jamming has a Gaussian tail  $p(f/p)\sim \exp(-A(f/p)^2)$, implying $\delta p\sim \frac{1}{\sqrt{\log{N}}}$, supposing again, that scaling holds till to the very first events, we obtain, using (\ref{eq:23}), the relation 
  \begin{eqnarray}
    \label{eq:24}
    p\sim\frac 1{\sqrt{|\ln((\Delta\sigma)|}}, 
  \end{eqnarray}
and by dimensional reasons, $e \sim\frac {\Delta\sigma}{\sqrt{|\ln((\Delta\sigma)|}}$. We remark, that these relations depend critically from the 
Gaussian tail of the distribution of the scaled forces at large argument. With a different tail this argument would give  a different dependence, e.g. a stretched exponential with stretching exponent $1/b$, would give rise to a log with a power $-b$ as in (\ref{eq:6}). It would be interesting to see 
if such a situation could produce in a physical system.

\section{Conclusions and perspectives}
We used the athermal adiabatic compression algorithm to surf on isostatic marginally stable minima that arise in the jammed phase of a prototype random continuous optimization problem, the spherical perceptron with linear cost function.
Using this algorithm we have studied the statistics of plastic events that arise when the system is compressed from the jamming transition inside the non-convex jammed phase. We found that inside the jammed phase,
such plastic events follow a statistics similar to hard sphere packings under quasi-static strain, which suggests that any local dynamics 
that destabilizes isostatic states and leads into other isostatic states has the same critical properties.  

Furthermore we have characterized the critical properties of the unjamming transition where there are logarithmic corrections to the scaling theory developed in \cite{OSLN03,GLS16}. The detailed form of these corrections are due to the Gaussian tail of the force distribution at jamming. 
In addition we have shown that the scaling of the number of overlaps as a function of the distance from jamming follows a power law behavior which is very different from what is found for harmonic spheres.

The approach we have developed suggests new ways to study landscapes with isostatic marginally stable minima. In our model it is possible to 
set up a landscape computation of the complexity of isostatic minima. Following \cite{BM80}, we can fix the control parameters $\alpha$ and $\sigma$  (or equivalently $p$) and count the number of solutions of the constitutive equations for local minima, see Eqs.~\eqref{grad_zero}, that respect the stability condition. In the same way fixing the isostatic index $C/N$ away from one,  we could count the solutions -if any- which are either hyperstatic or hypostatic. Note that hyperstatic (namely solutions with a number of contacts greater then $N$) solutions are likely to exist, see \cite{BIUWZ18, IUZ19}, but 
are expected to not be endowed with critical pseudogaps and cannot be dynamically reached\footnote{Unless some smart algorithms are employed and the nature of the problem is conserved only on average. In particular, following \cite{BIUWZ18, BLW18, IUZ19} we need to promote $\sigma$ to a fluctuating $\sigma_\mu$ which changes from gap to gap and impose that on average the statistics of $\sigma_\mu$ is conserved. This opens a new relaxing channel and therefore one can sensibly decrease the energy and increase the number of contacts.}.

On the numerical side, it will be interesting to use the algorithm to simulate compression-decompression cycles and see whether one can anneal  the system by a gentle driving. In this case it could be interesting to understand the span of the energy levels of local minima that can be achieved by using such algorithms as a function of the amplitude and speed of the compression cycles. It could well be that the dynamically accessible marginal manifolds at different energy levels are connected by small barriers, or at least barriers that scale as $N^t$ being $t$ an exponent that is $t<1$. 
This is for example the case of simple spin glasses as the Sherrington-Kirkpatrick model where $t=1/3$ \cite{VV89,KH91, BJ06}.

Finally we would like to comment more broadly on our findings.
We have used a quasistatic algorithm that explores isostatic minima of the linear perceptron. The present algorithm allows to 
study the statistics of the plastic events that follow a destabilizing perturbation. The minima that are explored appear to have the same 
universal marginal stability features of the ones found by the rather different gradient descent dynamics that we used in \cite{FSU19}. 
While the proof of such universality from the actual solution of the dynamics of specific algorithms may be probably achieved in some particular cases, notably the ones whose state evolution is described by the replica equations \cite{Mo19, EMS20} a generic picture of why such universality emerges beyond the algorithmic schemes employed (being gradient descent \cite{ABUZ18}, message passing or any local algorithm as in the present case) is partially lacking.
Indeed powerful local stability arguments \cite{LNSW10,LN10, MW15}, that are rather generic being independent on the algorithmic schemes, allow to obtain a set of non-trivial scaling relations between the critical exponents that describe isostatic points. However local stability alone does not lead to  a complete theory and probably it is not enough constraining to fix the value of the static critical exponents, nor the avalanche one $\tau$. 

In order to get these quantities one needs to do something more. The replica approach, provides a detailed scaling theory 
that predicts the numerical values of the exponents \cite{FSU19,CKPUZ14}. 
However, with replicas one either assumes Boltzmann sampling of the states or some large deviations \cite{Mo95,FP95} and therefore while providing a set of situations for which one can go beyond local stability and get critical exponents, it is still restrictive to those cases.

Seen this situation, we would like here to twist the usual perspective and propose a different point of view.
One of the outcomes of our analysis is that, in the jammed phase, the statistical properties of the avalanches
induced by a compression are essentially indistinguishable from those obtained from whatever random local perturbations that brings isostatic minima to isostatic minima with similar statistical properties. This is a property of \emph{stochastic stability}.  Stochastic stability, that has been studied extensively in the context of spin glasses \cite{MPV87}, is an important property of marginally stable disordered systems. Roughly speaking, it states that a small random perturbation leaves 
the relevant statistics of the system unchanged. In other words, random perturbations move the system to points that may be well far away in 
configuration space, but that have the same statistical properties of the old ones. Stochastic stability has been proven for the Gibbs measure in spin glasses \cite{AC98, Pa01, CG05, Co09}. At any finite temperature, when the system is at equilibrium, it implies that the static and dynamical responses are controlled by the 
overlap distribution that can be extracted from the equilibrium computation \cite{FMPP98,FMPP99}.  
A natural extension at zero temperature would imply that, starting from the ground state, the dynamic and static avalanches share the same statistics. 

A unifying way to address the universality properties of marginally stable minima could be to reverse the usual way of proceeding.
Rather than trying to analyze the asymptotic probability distribution of a given class of algorithms and show that it is stochastically stable, we may think to stochastic stability as a physical requirement
and characterize the invariant probability distributions over marginally stable states.
We emphasize that this way of treating the problem leads directly toward a statistical mechanics of marginally stable states.
In order to fix the ideas let us consider a modified Hamiltonian for our model
\begin{equation}
H_\Delta  = H + \frac{\Delta}{\sqrt N}  \z \cdot \x
\end{equation}
with $\z$ being a random vector whose components $Z_i$ are Gaussian with zero mean and unit variance. 
The perturbation is of relative order $1/N$ with respect to $H$ and therefore it does affect thermodynamic quantities (such as energy and pressure) only with subleading corrections in $N$. However it changes the equations for local minima. 
In particular the equations for forces gets shifted by a random term
\begin{equation}
\sum_{o} \frac{\xi_{o,i}}{\sqrt{N}} 
    +\sum_{c} f_{c} \frac{\xi_{c,i}}{\sqrt{N}}=
    \lm X_i + \frac{\Delta}{\sqrt N} Z_i\:.
\end{equation}
At this point we can start from a solution of the equilibrium equations at $\Delta=0$ and try to follow it when $\Delta$ is increased. 
The perturbation is much larger than $N^{-\beta}$ that triggers avalanches and therefore following it we will get to many other marginal states
far away from the original one. 
Stochastic stability states that the minima obtained by following such perturbations till to the larger distances have the very same properties 
of the starting one.  In order to look at the statistics of those minima, we could try to sample the random vector $\z$.
Very likely once the averages over $\z$ are taken, the dependence on the specific realization of the random patterns and the starting point $\x$, namely the particular solution at $\Delta=0$ are washed out, and the only leftover is in the energy level 
(and pressure and $\mu$) at which the original minimum is taken.
We believe that the replica scaling theory is a prominent candidate for the scale invariant distribution that could be obtained imposing stochastic stability.
One possibility is that averaging over $\z$ the universal part of the replica scaling solution should emerge.
We leave the analysis of this conjecture as a  program for  forthcoming work.

\section{Acknowledgements} 
SF and AS are supported by a grant from the Simons foundation (grant No. 454941, S. Franz). SF is a member of the Institut Universitaire de France.
This  work  was  supported  by  ``Investissements  d'Avenir"  LabEx-PALM  (ANR-10-LABX-0039-PALM).

\bibliography{HS}

\end{document}